%% file: em.tex
\documentclass{emulateapj}

\lefthead{VAN DER WEL ET AL.}
\righthead{EVOLUTION OF CLUSTER AND FIELD MDR}
\slugcomment{To Appear in ApJ, 670, 206}

\begin{document}

\title{The Evolution of the Field and Cluster Morphology-Density
Relation for Mass-Selected Samples of Galaxies}

\author{A. van der Wel\altaffilmark{1}, B. P. Holden\altaffilmark{2},
  M. Franx\altaffilmark{3}, G.D. Illingworth\altaffilmark{2},
  M. P. Postman\altaffilmark{4}, D. D. Kelson\altaffilmark{5}, \\
  I. Labb\'e\altaffilmark{5}, Wuyts, S.{3},
  J. P. Blakeslee\altaffilmark{6}, H. C. Ford\altaffilmark{1}}

\altaffiltext{1}{Department of Physics and Astronomy,
Johns Hopkins University, 3400 North Charles Street, Baltimore, MD
21218; wel@pha.jhu.edu}
\altaffiltext{2}{University of California Observatories/Lick Observatory, University of California, Santa Cruz, CA 95064}
\altaffiltext{3}{Leiden Observatory, Leiden University, P.O.Box 9513, NL-2300 AA Leiden, Netherlands}
\altaffiltext{4}{Space Telescope Science Institute, 3700 San Martin Drive, Baltimore, MD 21218}
\altaffiltext{5}{Carnegie Observatories, Carnegie Institution of Washington, 813 Santa Barbara Street, Pasadena, CA 91101}
\altaffiltext{6}{Department of Physics and Astronomy, Washington State University, Pullman, WA 99163-2814}
\altaffiltext{7}{Based
on observations with the \textit{Hubble Space Telescope}, obtained at
the Space Telescope Science Institute, which is operated by AURA,
Inc., under NASA contract NAS5-26555, and observations made with the
\textit{Spitzer Space Telescope}, which is operated by the Jet
Propulsion Laboratory, California Institute of Technology, under NASA
contract 1407.}

\begin{abstract}
  The Sloan Digital Sky Survey (SDSS) and photometric/spectroscopic
  surveys in the GOODS-South field (the Chandra Deep Field-South,
  CDF-S) are used to construct volume-limited, stellar mass-selected
  samples of galaxies at redshifts $0<z<1$.  The CDF-S sample at
  $0.6<z<1.0$ contains 207 galaxies complete down to $M=4\times
  10^{10}~M_{\odot}$ (for a ``diet'' Salpeter initial mass function),
  corresponding to a luminosity limit for red galaxies of
  $M_{\rm{B}}=-20.1$.  The SDSS sample at $0.020<z<0.045$ contains
  2003 galaxies down to the same mass limit, which corresponds to
  $M_{\rm{B}}=-19.3$ for red galaxies.  Morphologies are determined
  with an automated method, using the S\'ersic parameter $n$ and a
  measure of the residual from the model fits, called ``bumpiness,''
  to distinguish different morphologies.  These classifications are
  verified with visual classifications.  In agreement with previous
  studies, $65\%-70\%$ of the galaxies are located on the red
  sequence, both at $z\sim 0.03$ and at $z\sim 0.8$. Similarly,
  $65\%-70\%$ of the galaxies have $n>2.5$. The fraction of E+S0
  galaxies is $43\%\pm 3\%$ at $z\sim 0.03$ and $48\%\pm 7\%$ at
  $z\sim 0.8$; i.e., it has not changed significantly since $z\sim
  0.8$.  When combined with recent results for cluster galaxies in the
  same redshift range, we find that the morphology-density relation
  for galaxies more massive than $0.5M^*$ has remained constant since
  at least $z\sim 0.8$. This implies that galaxies evolve in mass,
  morphology, and density such that the morphology-density relation
  does not change.  In particular, the decline of star formation
  activity and the accompanying increase in the stellar mass density
  of red galaxies since $z\sim 1$ must happen without large changes in
  the early-type galaxy fraction in a given environment.
\end{abstract}


\section{INTRODUCTION}

The morphology-density relation \citep[MDR;][]{dressler80}, i.e., the
observation that high-density environments contain a higher fraction
of early-type galaxies than low-density environments, provides a clue
about the formation and evolution of galaxies, and suggests that
environment plays an important role in shaping the galaxy population.
The interpretation of the MDR, however, is not straightforward because
of additional correlations between morphology, galaxy mass, color,
star formation history, and metallicity; all quantities that depend on
environment
\citep[e.g.,][]{hogg03,kauffmann03a,kauffmann04,blanton05a,baldry06}.
There are strong indications that the stellar mass of a galaxy,
independently of the environment, determines the color
\citep{baldry06} and concentration \citep{kauffmann04} of a galaxy. In
addition, the environment plays a role in star formation activity
\citep{kauffmann04}.

It is a matter of debate whether the fate of a galaxy is ultimately
determined by its mass or its environment. Studies of galaxy
properties such as those mentioned above at higher redshifts help to
answer those questions, and there is abundant evidence for both
scenarios. The massive galaxy populations of $z\sim 1$ clusters appear
to be fully assembled and passively evolving
\citep[e.g.,][]{depropris07}, whereas the galaxy population as a whole
undergoes significant evolution between $z=1$ and the present, both in
terms of mass and color \citep[e.g.,][]{bell04b} and in terms of star
formation \citep[e.g.,][]{lefloch05}. This points to a strong relation
between formation epoch and environment. On the other hand, high-mass
galaxies have had lower specific star formation rates than low-mass
galaxies since at least $z\sim 2$ \citep[e.g.,][]{juneau05,noeske07},
and there is only a very modest age difference between field and
cluster early-type galaxies with masses $M>10^{11}~M_{\odot}$ at
$z\sim 1$ \citep{vandokkum07}, with possibly a stronger age dependence
on mass than environment \citep{vanderwel05,treu05b}. All these
results, at first sight contradictory, will have to be reconciled with
each other.

The MDR, and its evolution with redshift, plays a crucial role in our
understanding of the effect of the environment on galaxy evolution,
and significant evolution of the early-type galaxy fraction has been
found between $z\sim 1$ and the present
\citep[e.g.,][]{dressler97,smith05,postman05}.  These studies are
based on samples that are selected by luminosity, and to ensure that
the $z\sim 1$ samples contain the progenitors of the galaxies in the
local samples, the luminosity limit for samples at different redshifts
is corrected for evolution. It is assumed that the evolution is the
same for all galaxies, which might not be the case. In addition,
luminosity is very sensitive to bursts of star formation that are
likely far more prevalent at $z\sim 1$ than in the local
universe. Therefore, it is worthwhile to adopt an alternative
selection method that is less sensitive to potentially rapidly
changing galaxy properties such as the star formation rate.  Stellar
mass is a quantity that changes less rapidly than luminosity, such
that it might be a more suitable tracer of the $z\sim 1$ progenitors
of local galaxies.  In addition, it is more straightforward to compare
stellar masses to model predictions.  Obviously, mass is not an ideal
tracer of the evolving galaxy population, just like luminosity, as it
increases through star formation and mergers. Still, in \citet[][,
hereafter Paper I]{holden07} we show that there is no discernible
evolution between $z=0.83$ and the present in the morphological mix of
galaxies that are more massive than $4\times 10^{10}~M_{\odot}$
($\sim$25\% of the mass of a typical cluster galaxy, $\sim$50\% of the
mass of a typical field galaxy).  The evolution in the early-type
fraction seen in luminosity-selected samples of cluster galaxies is
due to the relative increase in the number of blue, star-forming,
low-mass galaxies.

Given the abundant evidence for significant evolution since $z\sim 1$
in the galaxy population outside clusters, it is important to measure
the early-type galaxy fraction at lower densities.  It is not
inconceivable that the early-type galaxy fraction does evolve strongly
in those environments because of the strong evolution of the galaxy
population in the field and in groups. On the other hand, if
morphological evolution is primarily driven by environmental effects,
the morphological mix of the galaxy population in a given density
might not evolve. Several authors have studied the morphological mix
of galaxies at $z\sim 1$
\citep[e.g.,][]{bell04a,bundy05,capak07,abraham07}. A range of
techniques and sample selection methods is used in those
investigations, and it is not always clear how the high-$z$ results
compare to local studies and how sample selection affects the
interpretation. In particular, mass estimates and morphologies are
generally not obtained with the same method for local and distant
galaxy samples, such that systematic errors may contribute to the
observed evolution in, for example, the early-type fraction.

In this paper we construct mass-selected samples of low- and
high-redshift galaxies with morphologies classified in an internally
consistent manner, such that the morphological mix of the galaxy
population at low and high redshift can be directly compared both with
each other and with our recent results on samples of cluster galaxies
over the same redshift range (Paper I). In \S~2 we describe our
galaxy samples extracted from the Sloan Digital Sky Survey (SDSS) at
redshifts $0.020<z<0.045$ and the Chandra Deep Field-South (CDF-S) at
redshifts $0.6<z<1.0$ and determine galaxy masses and morphologies.
In \S~3 we present and discuss our results and their implications
for the evolution of the early-type galaxy fraction in our field
samples and, through comparison with cluster galaxies, how the MDR
evolves with redshift. Finally, in \S~4 we put our results in the
context of the evolving red sequence and the decline in the average
cosmic star formation rate density.  We adopt the standard
cosmological parameters, $(\Omega_{\rm{M}},~\Omega_{\Lambda},~h) =
(0.3,~0.7,~0.7)$.

\section{DATA}

First, we construct a complete, stellar-mass-selected, volume-limited
sample of galaxies at redshifts $0.02<z<0.045$ from the Sloan Digital
Sky Survey \citep[SDSS;][]{york00}, Data Release 5
\citep[DR5;][]{adelman07}.  Similarly, we construct a stellar
mass-selected, volume-limited sample at $0.6<z<1.0$ in the Chandra
Deep Field-South (CDF-S).  Then we determine the morphologies of the
galaxies in both samples in a consistent manner and quantify the
environment (parameterized by the local surface galaxy density) in
which the galaxies are situated.

\subsection{A Mass-Selected, Volume-Limited SDSS Galaxy Sample}\label{secsdss}

The sample consists of two redshift bins: galaxies at redshifts
$0.02<z<0.03$ with stellar masses $4\times
10^{10}<M/M_{\odot}<1.6\times 10^{11}$, and galaxies at redshifts
$0.035<z<0.045$ with stellar masses $M > 1.25\times
10^{11}~M_{\odot}$.  The reason for this split is to ensure
spectroscopic completeness on both the faint and the bright end and to
minimize systematic problems with the photometry of bright galaxies
(see below). The $g$-band magnitudes of the galaxies in these samples
range from $g_{\rm{mod}}=13.7$ to 16.5. Here $g_{\rm{mod}}$ is the
model magnitude from the SDSS pipeline (see below for the
definition). The upper limit $g_{\rm{mod}}=13.7$ ensures that this
sample is not incomplete due to the cutoff at bright magnitudes in the
SDSS spectroscopic survey \citep{strauss02}.  This source of
incompleteness is essentially a surface brightness limit imposed to
avoid saturation.  The lower magnitude limit $g_{\rm{mod}}=16.5$
ensures spectroscopic completeness at the faint end \citep{blanton05b}
and the feasibility of determining morphologies without being limited
by background noise.  Finally, only those galaxies located at least
130 pixels from the edge of the image tiles are included in the
sample, in order to avoid problems with the morphological analysis
described below. This does not introduce systematic effects other than
decreasing the volume of the sample by 28\%.  The lower mass limit of
$M=4\times 10^{10}~M_{\odot}$ is chosen to match the $z\sim 0.8$
sample that we construct below (see \S~\ref{secdatacdfs}).
Approximately two-thirds of the stars are located in galaxies with
$M>4\times 10^{10}~M_{\odot}$ \citep{bell03}, such that in terms of
stellar mass our samples contain a representative fraction of the
galaxy population.

\input{tab1.tex}

Total magnitudes can be underestimated due to overestimation of the
sky brightness by the SDSS reduction pipeline \citep{lauer07}. The
magnitude of this effect depends on both galaxy size and surface
brightness, i.e., to first order on apparent magnitude.  Since we are
working with a large sample we are primarily interested in a
statistical, magnitude-dependent correction for the sample as a whole.
For a dozen galaxies with magnitudes across the range of our sample we
determine total magnitudes by hand.  The true background is determined
by SExtractor \citep{bertin96}, using the global background mode, and
the total flux is measured with the {\tt ellipse} task in IRAF.  We
find that $g_{\rm{tot}}=g_{\rm{mod}}-0.2(15-g_{\rm{mod}})$ for
$g_{\rm{mod}}<15$ and $g_{\rm{tot}}=g_{\rm{mod}}$ for
$g_{\rm{mod}}>15$.  This correction is of similar magnitude to the
corrections applied by \citet{vonderlinden07}, who designed a method
to obtain accurate, corrected magnitudes for individual galaxies.  Our
correction is not perfect for individual galaxies but does provide a
sufficiently accurate correction for the sample as a whole. The impact
of the correction on the stellar masses that we derive below is $\sim$
0.1 dex for the brightest galaxies, such that even if the correction
we apply is uncertain on the 50\% level, the effect on the stellar
masses is only 0.05 dex, which is negligible with respect to the other
uncertainties on the mass estimates.  The systematic error will be
still lower.

We adopt $u-g=u_{\rm{mod,c}}-g_{\rm{mod,c}}$ and
$g-r=g_{\rm{mod,c}}-r_{\rm{mod,c}}$ as the galaxy colors, where
$u_{\rm{mod,c}}$, $g_{\rm{mod,c}}$, and $r_{\rm{mod,c}}$ are the model
magnitudes, corrected for Galactic extinction. Here $r_{\rm{mod,c}}$
is either the De Vaucouleurs model magnitude or the exponential model
magnitude, depending on which one fits best to the observed surface
brightness profile in the $r$ band.  The best-fit radius, ellipticity,
and orientation as determined in the $r$ band are subsequently used to
measure $u_{\rm{mod,c}}$ and $g_{\rm{mod,c}}$, with only the surface
brightness as a free parameter \citep[see][for a description of the
data reduction pipeline]{lupton01}.  The colors thus obtained are
therefore equivalent to the color within the $r$-band half-light
radius.

$K$-corrections are not negligible, even at such low redshifts
\citep{blanton03}.  We derive $g_{\rm{tot},0}$, $(u-g)_0$, and
$(g-r)_0$ ($g$-band magnitudes, and $u-g$ and $g-r$ colors
$K$-corrected to $z=0$) by following the technique applied to higher
redshifts galaxies by \citet{blakeslee06}, \citet{holden06} and Paper
I.  We derive linear relations between colors and magnitudes in the
observed and rest frames from synthetic spectra for stellar
populations from \citet{bruzual03} with a range of metallicities and
star formation histories (exponentially declining star formation rates
with different timescales).  In Table \ref{tab1} we show the
transformations for redshifts $z=0.02$, 0.03, and 0.04.

Subsequently, stellar masses are estimated using the relation between
$(g-r)_0$ and $M/L$ derived by \citet{bell03} for a ``diet'' Salpeter
initial stellar mass function (IMF):
\begin{eqnarray*}
\log \bigg ( \frac{M}{L_{\rm{g}}} \bigg ) = 1.519 (g-r)_0 - 0.499 .
\end{eqnarray*}
The diet Salpeter IMF \citep[e.g.,][]{bell03} gives values of $M/L$
that are 0.15 dex lower than in the standard Salpeter IMF, which is
due to the flat slope for stellar masses below $0.35~M_{\odot}$.  The
$K$-corrections and color-$M/L$ conversions can be combined to give a
unique relation between stellar mass, apparent magnitude, color, and
redshift:
\begin{eqnarray*}
\log\bigg (\frac{M}{M_{\odot}}\bigg ) = f(g_{\rm{tot}};(g-r);z) = \\
-0.399 \bigg [g_{\rm{tot}} - 3.65 (g-r) - 4.79 \log\bigg (
  \frac{z}{0.03}\bigg ) - 39.55 \bigg].
\end{eqnarray*}
In order to compare with previous work and the $z\sim 0.8$ sample
described below, we derive the standard $U-B$ and $B-V$ colors and
$B$-band magnitudes in the Vega system.

\begin{figure}
\epsscale{1.2}
\plotone{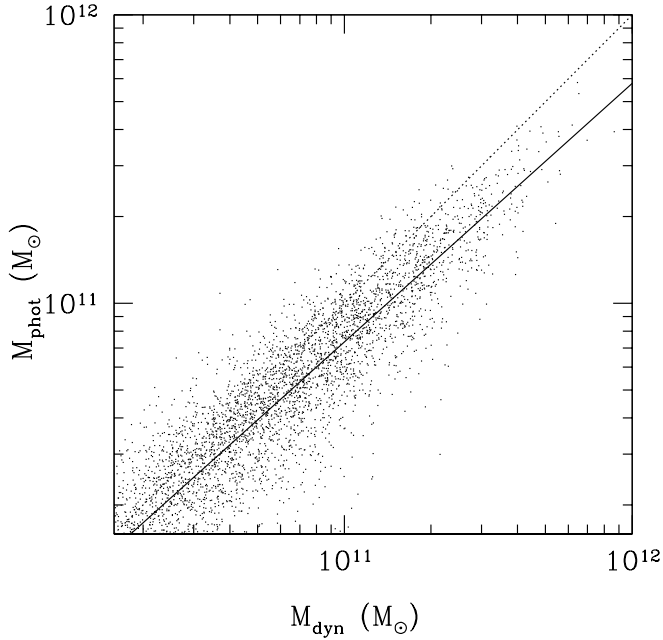}
\caption {Comparison between stellar masses (as used in this paper)
 and kinematic masses ($\propto R\sigma^2$) for galaxies at redshifts
 $0.030<z<0.045$. The dotted line indicates the one-to-one
 correspondence between $M_{\rm{dyn}}$ and $M_{\rm{phot}}$. The solid
 line is the best linear fit, which has slope 0.9 and shows that the
 offset between the two mass estimates is, on average, 0.10 dex.  The
 scatter is 0.17 dex.}
\label{masscheck_sdss}
\end{figure}

In Figure \ref{masscheck_sdss} we compare our stellar mass estimates
with kinematic mass estimates, where the latter is defined as
$M_{\rm{dyn}}/M_{\odot}=\log(R)+2\log(\sigma_c)+6.07$
\citep{jorgensen96}.  Here, $R$ is either the De Vaucouleurs or the
exponential disk scale radius in kiloparsecs as measured in the $g$
band, depending on which profile provides the better fit, and
$\sigma_c$ is the aperture-corrected velocity dispersion
\citep[see][]{bernardi03a} in kilometers per second. There is an
average, systematic offset of 0.10 dex, the stellar mass estimates
being lower than the kinematic mass estimates. This can be due to the
choice of the diet Salpeter IMF and/or the presence of nonbaryonic
matter, even though $\sigma$ is measured at the centers of the
galaxies, which are dominated by stars.  The offset varies slightly
with mass (by $\lesssim \pm 0.05$ dex) for galaxies with
$M_{\rm{phot}}>4\times 10^{10}~M_{\odot}$, the mass limit of the
sample used in this paper.  Such a mass dependence of the difference
between kinematic and stellar mass estimates is also seen at higher
redshifts (see \S~\ref{secdatacdfs}). The scatter in Figure
\ref{masscheck_sdss} is 0.17 dex, which implies that the errors in the
photometric mass estimates for individual galaxies are less than
$\sim$50\%. Note that this is an upper limit on the real uncertainty
since kinematic mass estimates have their own uncertainties. We
conclude that the color measurements and the resulting stellar mass
estimates are sufficiently accurate for our purposes.

The final sample consists of 2003 galaxies that are more massive than
$M=4\times 10^{10}~M_{\odot}$. Of those, 545 are in the
$0.035<z<0.045$ subsample of galaxies with masses $M>1.25\times
10^{11}~M_{\odot}$. When we compute numbers and fractions of numbers
of galaxies we give those high-mass galaxies lower weight (0.395),
which is the ratio of the volumes occupied by the $0.035<z<0.045$
sample and the $0.02<z<0.03$ sample.

\begin{figure}
\epsscale{1.2}
\plotone{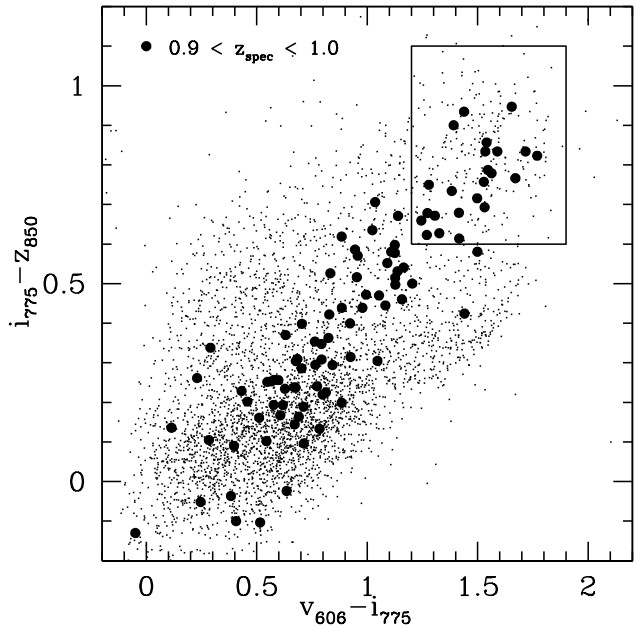}
\caption{ \textit{Small circles:} Galaxies brighter than $z_{850}=24$
  in the CDF-S. \textit{Large circles:} Galaxies with spectroscopic
  redshifts $0.9<z<1.0$.  A mass-selected sample will be biased most
  strongly at the high-redshift end ($z=1$) and against those galaxies
  with the highest $M/L$, i.e., those with the reddest colors.  The
  area indicated by the rectangle indicates the colors of the reddest
  galaxies at $z\sim 1$.  In Fig. \ref{compl} we address the redshift
  completeness of galaxies with such colors.}
\label{vi_iz}
\end{figure}

\subsection{A Mass-Selected, Volume-Limited Galaxy Sample in the CDF-S}\label{secdatacdfs}

We combine photometric and spectroscopic redshift samples to construct
a stellar mass-limited sample of galaxies in the CDF-S in the redshift
range $0.6<z<1.0$. We use the photometric redshift catalog by
\citet{wuyts07}, which is based on a $K$-band-selected sample.  We
supplement these with spectroscopic redshifts from the literature
\citep{lefevre04,mignoli05,vanderwel05,vanzella06}. Only those
galaxies for which the spectroscopic redshifts have been measured with
reasonable confidence (best or second-best quality flags) are
considered. Of the final sample, 142 out of 207 or 69\% have
spectroscopic redshifts. We refer to the respective publications for
explanations of the confidence indicators.

Extensive comparisons by Wuyts et al. show that in the redshift range
of our sample, there are no systematic differences between
spectroscopic and photometric redshifts, regardless of color and
magnitude cuts imposed on the sample.  There are occasional
catastrophic failures, but their number is small ($1\%-2\%$), and
therefore this does not introduce significant systematic effects.

The $K$-band limit of the photometric redshift catalog determines the
mass completeness limit of our sample. Since we use $v_{606}$-,
$i_{775}$-, and $z_{850}$-band ACS (\textit{Hubble Space Telescope}
Advanced Camera for Surveys) photometry to calculate masses,
luminosities, and rest-frame colors we quantify the completeness limit
in terms of $z_{850}$-band total magnitudes.  The $z_{850}$-band total
magnitudes are the SExtractor Best magnitude, but we add flux (0.2
mag) to correct for light outside the aperture
\citep{benitez04,blakeslee06}.  We measure $v_{606}-i_{775}$ and
$i_{775}-z_{850}$ within a $0.5''$ diameter aperture, and we correct
for differential point-spread function (PSF) effects. The latter
correction is negligible in $v_{606}-i_{775}$ but ranges up to 0.07
mag (for point sources) in $i_{775}-z_{850}$. The aperture is chosen
such that we measure the color within the effective radius or
half-light radius of a typical $z\sim 0.8$ galaxy, which corresponds
closely to the model colors we use to define the colors of the
galaxies in the $z\sim 0.03$ sample (see \S~\ref{secsdss}).

\begin{figure}
\epsscale{1.2}
\plotone{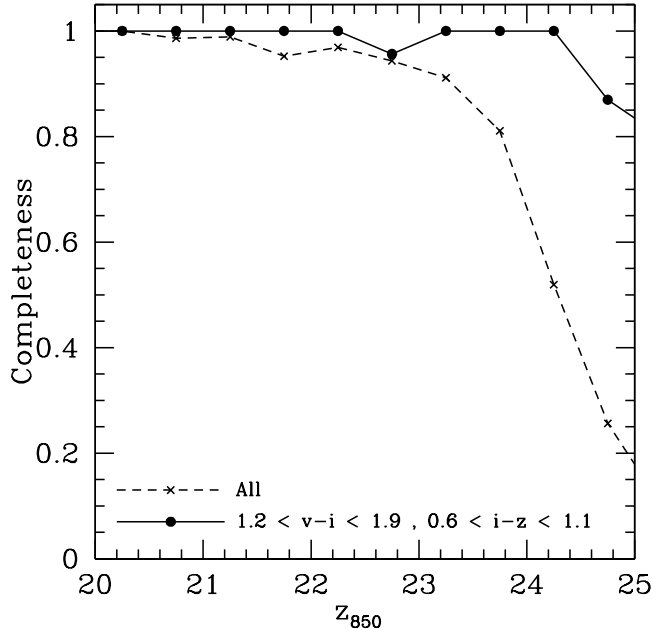}
\caption{ \textit{Dashed line with crosses:} Fraction of galaxies in
  the CDF-S with known spectroscopic and/or photometric redshifts as a
  function of $z_{850}$ magnitude.  The completeness begins to
  decrease at $z_{850}\sim 23-23.5$.  \textit{Dotted line with
  circles:} Completeness of galaxies within the rectangle indicated in
  Fig. \ref{vi_iz}, i.e., those galaxies with colors similar to the
  reddest galaxies at $z\sim 1$.  This sample is essentially complete
  down to $z_{850}=24$, which implies that we can use $z_{850}=24$ to
  define the mass limit of the sample.}
\label{compl}
\end{figure}

Selection effects in a mass-limited sample will be strongest at the
high end of a redshift bin (here $0.6<z<1.0$) and for the galaxies
with the highest mass-to-light ratio ($M/L$), i.e., the reddest
galaxies.  In Figure \ref{vi_iz} we show the $v_{606}-i_{775}$ and
$i_{775}-z_{850}$ colors of the galaxies in the CDF-S with
spectroscopic redshifts $0.9<z_{\rm{spec}}<1.0$.  The rectangle, which
outlines the region with the reddest galaxies at that redshift,
indicates for which colors incompleteness first begins to affect our
mass-limited sample.  In Figure \ref{compl} we show that for those
colors the $K$-band-selected photometric redshift catalog is virtually
complete down to $z_{850}=24$, fainter than the completeness limit of
galaxies with a typical color. In the following, we adopt $z_{850}=24$
as the magnitude limit which we use to define the mass completeness
limit of our sample.  Bluer galaxies near this mass limit are much
brighter, and incompleteness does not play a role for those objects.
The reason for using a $z_{850}$-band magnitude limit to define the
mass completeness limit instead of the $K$-band magnitude limit of the
photometric redshift catalog is that we derive stellar masses by
estimating mass-to-light ratios in the rest-frame $B$-band, which
corresponds to the observed $z$ band at $0.6<z<1.0$.

Stellar masses are derived with the same method as used for the $z\sim
0.03$ galaxy sample. First, observed magnitudes and colors ($z_{850}$,
$v_{606}-i_{775}$, and $i_{775}-z_{850}$) are used to derive
rest-frame $B$-band luminosities and rest-frame $U-B$ and $B-V$
colors.  For $z<0.85$ we use $v_{606}-i_{775}$ to derive $(U-B)_0$ and
$v_{606}-z_{850}$ to derive $(B-V)_0$ and $M_{\rm{B},0}$; for $z>0.85$
we use $v_{606}-z_{850}$ to derive $(U-B)_0$ and $i_{775}-z_{850}$ to
derive $(B-V)_0$ and $M_{\rm{B},0}$.  These choices optimally match
the observed and rest-frame bands, and the derived rest-frame colors
are well behaved as a function of redshift. Examples for various
redshifts are given in Table \ref{tab1}.  The rest-frame $U$-, $B$-,
and $V$-band magnitudes and colors are in the Vega system, as opposed
to the other magnitudes and colors, which are in the AB system. This
is necessary to facilitate the comparison with earlier results.
Finally, the empirical relation between $M/L$ and $B-V$ from
\citet{bell03} is used to estimate the stellar masses:
\begin{eqnarray*}
\log\bigg ( \frac{M}{L_{\rm{B}}}\bigg )) = 1.737 (B-V)_0 - 0.994.
\end{eqnarray*}

For 14 early-type galaxies in the $z\sim 0.8$ sample kinematic
measurements are available from \citet{vanderwel05}. There is a
systematic offset between the kinematic and stellar mass estimates of
0.1 dex (the kinematic masses being larger) and a scatter of 0.3 dex.
These results are consistent with earlier determinations of the
robustness of stellar mass estimates at $z\sim 0.8-1$
\citep{vanderwel06b,holden06}.  Most importantly, the systematic
difference between kinematic and stellar mass estimates is the same
(0.1 dex) for the $z\sim 0.03$ sample and the $z\sim 0.8$ sample.
Therefore, we conclude that our stellar mass estimates do not
introduce systematic effects into our samples that exceed 0.1 dex in
mass as was found earlier by both \citet{vanderwel06b} and
\citet{holden06}. Uncertainties at that level do not significantly
affect our results.  The difference between the kinematic and stellar
mass estimates is mass dependent \citep[see, e.g.,][and Paper
I]{holden06}, but since this is also the case for our $z\sim 0.03$
sample (see \S~\ref{secsdss}), and at the same level, this does not
affect our results significantly.

\begin{figure}
\epsscale{1.2}
\plotone{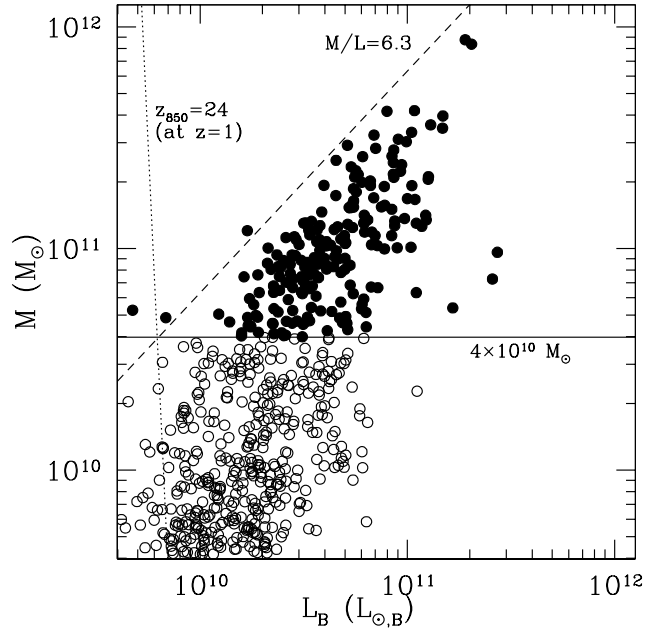}
\caption{ Rest-frame $B$-band luminosity vs. stellar mass for galaxies
  in the CDF-S with redshifts $0.6<z<1.0$, using spectroscopic
  redshifts if available, and, to ensure completeness, including
  galaxies with only photometric redshifts as well.  Our magnitude
  completeness limit $z_{850}=24$ corresponds to the luminosity limit
  at $z=1$ (\textit{dotted line}).  In addition, we adopt
  $\log(M/L_B)=0.8$ (\textit{dashed line}) as the maximum value that
  the $M/L$ of any galaxy can take.  This is a reasonable assumption
  as this is the highest $M/L$ for galaxies at the high-mass,
  high-luminosity end of our sample.  The magnitude and $M/L$ limits
  cross at $M>4\times 10^{10}~M_{\odot}$.  This is the mass
  completeness limit of our sample. At lower masses, our sample is
  incomplete for faint, red (high $M/L$) galaxies (see Fig.
  \ref{compl}).  The mass-selected sample contains 207 galaxies, 142
  of which have spectroscopic redshifts.}
\label{L_M}
\end{figure}

\begin{figure*}
\epsscale{1.1}
\plottwo{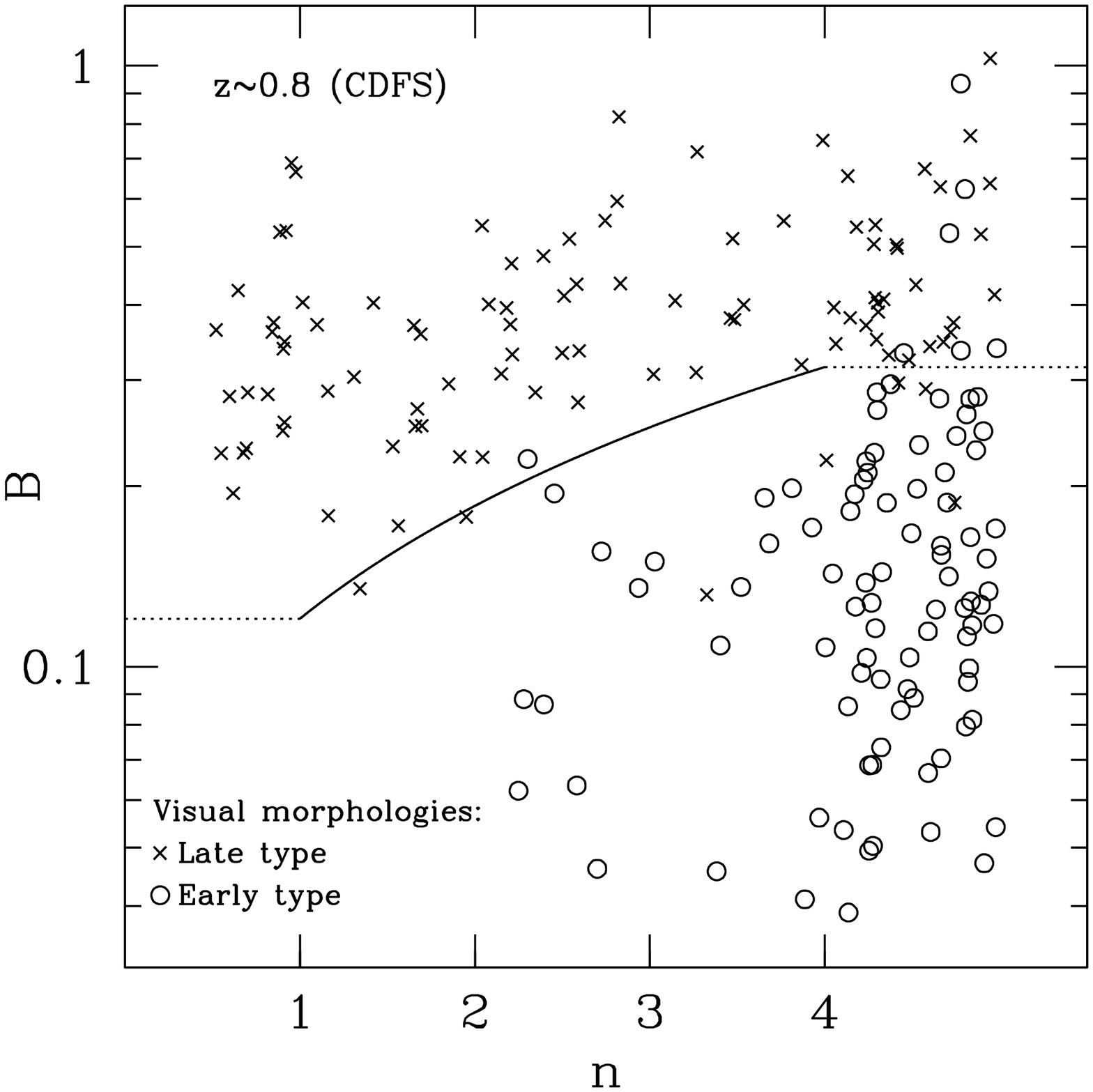}{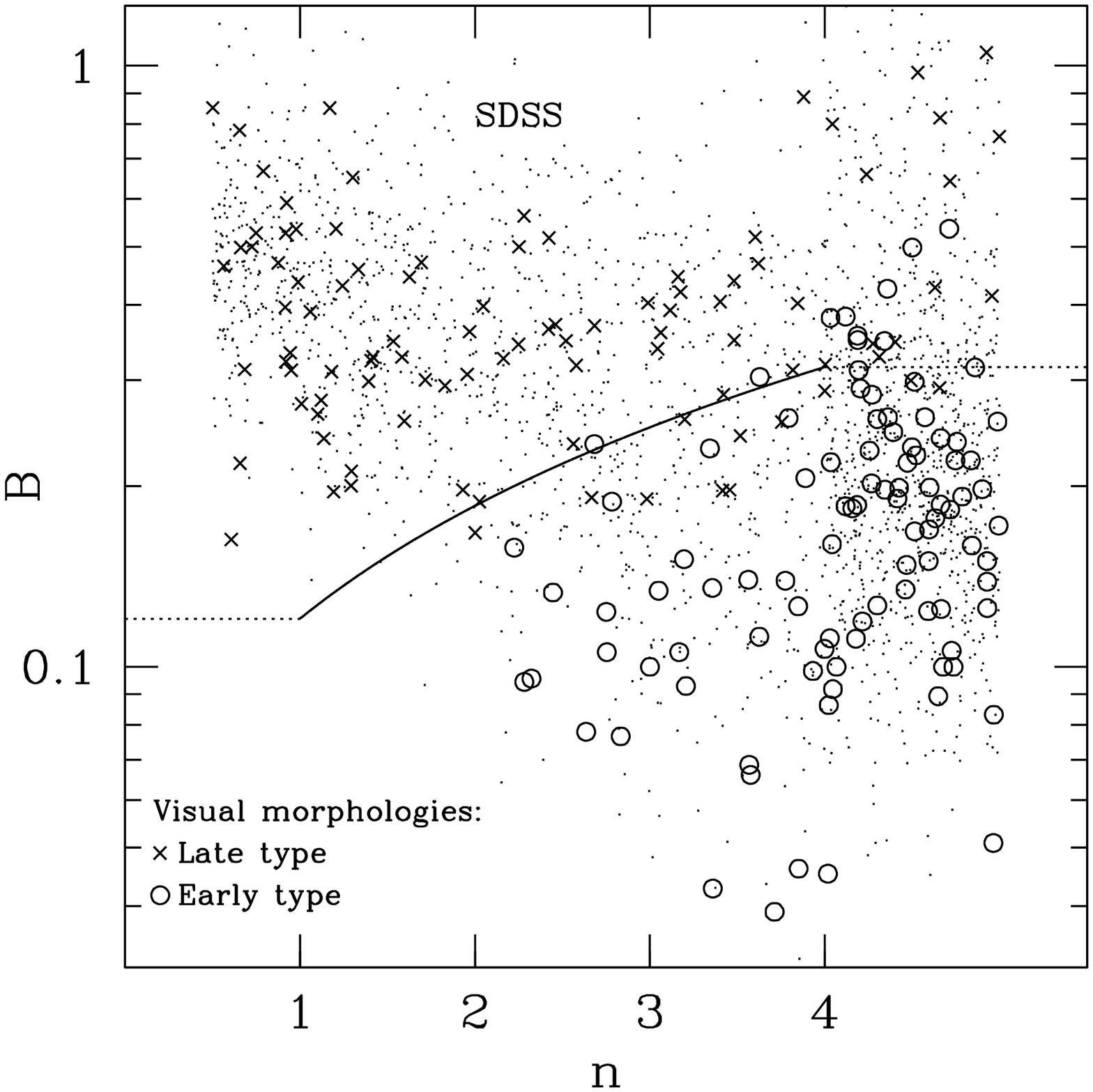}
\caption{ S\'ersic $n$ vs. bumpiness $B$ for all 207 galaxies in the
  mass-selected $z\sim 0.8$ sample (\textit{left}) and the 2003
  galaxies in the $z\sim 0.03$ sample (\textit{right}). The 200
  galaxies with visual classifications are separately indicated, and
  the remainder of the sample is indicated with small dots.  The
  S\'ersic index $n$ is restricted to be between $n=1$ and 4, but to
  enhance the readability of the figure we have randomly distributed
  the values of $n$ between 0.5 and 1 for galaxies with $n=1$ and
  between $4<n<5$ for galaxies with $n=4$. Visual classifications are
  indicated by different symbols: early types are indicated by open
  circles, late types by crosses.  The curved line, $B=0.065(n+0.85)$,
  indicates the separation between early types and late types
  according to the $B-n$ classification.  The correspondence between
  the visual and automated classifications is very good. See text for
  more details and Figure \ref{panels} for illustration.}
\label{morphcheck}
\end{figure*}

For late-type galaxies a comparison between kinematic and stellar mass
estimates is less straightforward. Recently, \citet{kassin07} showed
that the relation between stellar mass and kinematic mass as defined
by a combination of the rotation speed and the velocity dispersion of
the gas does not vary by more than $\sim$0.1 dex over the redshift
range $0.1<z<1.2$.  However, the stellar masses from Kassin et al.
are determined by fitting stellar population models to broadband
spectral energy distributions, which is not directly comparable with
our method.  If Kassin et al. had used the color-$M/L$ relations from
\citet{bell03}, the stellar masses would have changed such that at
$z\sim 1$ they would be $\sim$0.2 dex higher than those of local
galaxies with the same kinematic mass (Paper I). We comment on the
effects of possible biases where relevant.

In Figure \ref{L_M} we show the rest-frame $B$-band luminosities and
stellar masses of the galaxies in our $z\sim 0.8$ sample.  The highest
$M/L$ that occurs (for luminous, red galaxies) is
$M/L_{\rm{B}}=6.3$. We adopt this as the $M/L$ upper boundary, which,
combined with the magnitude limit derived above ($z_{850}=24$), gives
us the mass completeness limit of our sample: $M=4\times
10^{10}~M_{\odot}$ (see Fig. \ref{L_M}).  The mass-limited sample
contains 207 galaxies.

The average $M/L$ is $M/L_{\rm{B}}=2.4$ for the galaxies with
$M>4\times 10^{10}~M_{\odot}$. For the $z\sim 0.03$ sample this is
$M/L_{\rm{B}}=4.9$. This implies 0.75 mag evolution between the two
samples in the $B$-band, or 1 mag per unit redshift. This is
consistent with the evolution of $L^*$ \citep[e.g.,][]{brown07}, but
somewhat less than the $M/L_{\rm{B}}$ evolution inferred via
fundamental plane analyses for $z\sim 1$ field early-type galaxies
with similar masses \citep[0.55
dex;][]{vanderwel05,treu05b,vandokkum07}.  This difference may contain
contributions from a systematic uncertainty in the high-$z$ mass
estimates as described above, and an increased number of very red,
dusty galaxies at $z\sim 1$ with high $M/L$. 

\subsection{Morphological Classifications}\label{secmorph}

Traditionally, visual morphological classifications are used for
morphological studies of high-redshift galaxy populations, for example
the evolution of the MDR \citep[e.g.,][]{postman05}.  In recent years,
however, increased sample sizes have initiated the emergence of
several alternative, automated classification schemes
\citep{conselice00,abraham03,lotz04,blakeslee06}. In addition to the
traditional, visual classifications we therefore also deploy a
quantitative measure of the galaxy morphologies. This enables us to
analyze the large sample of $z\sim 0.03$ galaxies and to examine, in a
quantitative manner, the effect of cosmological surface brightness
dimming on the detectability of disks, spiral arms, and
irregularities. The visual classifications are used to verify the
robustness of this automated classification method.  In this paper we
simply distinguish between early- and late-type galaxies, or, in the
$T$-type classification scheme \citep[see][]{postman05}, between
galaxies with type $T\leq 0$ (E, E/S0, and S0) and galaxies with type
$T>0$ (Sa and later).  The separation into subclasses, for example E
and S0, will be investigated in a future paper.

Visual morphologies were determined by A. v. d. W.  by examining the
$g$-band images of 200 randomly chosen galaxies in the mass-selected
$z\sim 0.03$ sample and the $z_{850}$-band images of all 207 galaxies
in the $z\sim 0.8$ sample.  For 35 galaxies in the $z\sim 0.8$ sample
with $T\sim 0$ or peculiar morphologies M. P. also examined the images
and assigned morphologies.  This provides an estimate of the random
uncertainty and ensures that the same threshold ($T=0$) is adopted as
in \citet{postman05}.  Based on these results, we conservatively
assume a 5.5\% uncertainty in the early-type galaxy fraction due to
classification errors.  We note that the $z_{850}$-band imaging with
an integration time of 12,000 s and a $10\sigma$ point-source
detection limit of $z_{850}=27$ is very deep with respect to the
faintest galaxies in our sample with $z_{850}=24$.  In Figure
\ref{morphcheck} we indicate early-type galaxies with open circles and
late-type galaxies with crosses.

Quantitative morphologies are determined for all galaxies with the
technique outlined by \citet{blakeslee06}, who showed that the
S\'ersic parameter $n$ and bumpiness parameter $B$ can effectively
distinguish between early- and late-type galaxies.  GALFIT
\citep{peng02} is used to fit S\'ersic profiles to the $g$-band and
$z_{850}$-band images; $n$ is constrained to values between $n=1$ and
4. The only reason for not allowing larger values of $n$ is that the
effective radii for high values of $n$ become large and uncertain.
This would strongly affect the measurement of the bumpiness $B$, which
is the rms of the residual as measured within two effective radii.  In
order to reduce the effect of shot noise the residual is slightly
smoothed. In addition, other objects and a circular region around the
center are masked. This central region is masked out because of slight
PSF mismatches and because of central, small-scale deviations from a
S\'ersic profile. Because of the latter effect the adopted radius is
distance dependent. On visual inspection we choose a 4 pixel radius
for galaxies at $z=0.03$ in the low-$z$ sample and a 2 pixel radius
for galaxies at $z=0.80$ in the high-$z$ sample, both scaled inversely
proportional with angular diameter distance for galaxies at different
redshifts.

\begin{figure*}
\epsscale{2.4}
\plottwo{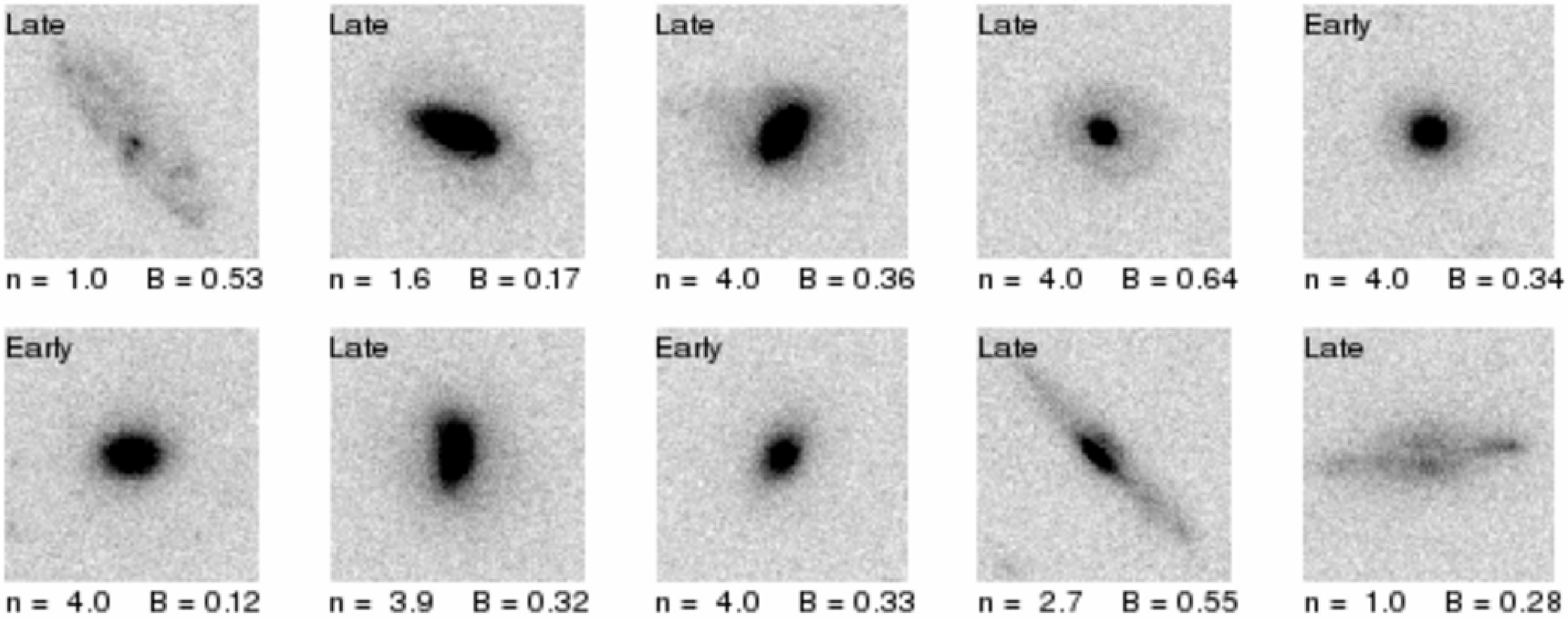}{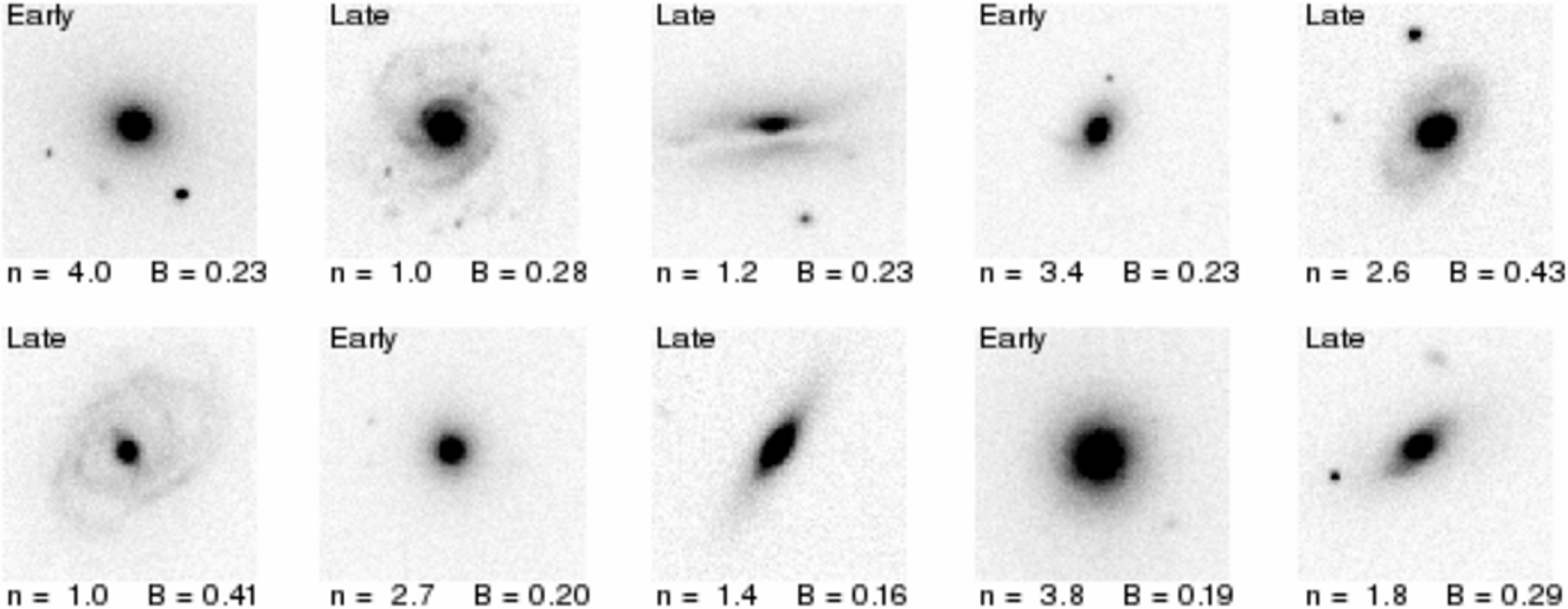}
\caption{\textit{Top two rows:} $3.4'' \times 3.4''$ $z_{850}$-band
 cutouts for 10 randomly chosen galaxies from the sample at
 $0.6<z<1.0$. \textit{Bottom two rows:} $51'' \times 51''$ $g$-band
 cutouts for 10 randomly chosen galaxies from the sample at $z\sim
 0.03$. The physical scale of both sets of images is comparable:
 $\sim$30 kpc. Visual classifications (late or early) and $n$ and $B$
 parameters are listed for each galaxy to illustrate the
 correspondence between the visual morphologies and the automated
 classification method (see also Fig. \ref{morphcheck}).}
\label{panels}
\end{figure*}

The advantage of the $B-n$ classification over other automated
classifiers is that PSF smearing is taken into account. This is
essential, in particular when redshift-dependent trends are
investigated in data sets with very different photometric
properties. In Figure \ref{panels} we show image cutouts and give the
$n$ and $B$ parameters for 20 randomly selected galaxies from our
$z\sim 0.0$ and 0.8 samples, in order to illustrate the correspondence
between the automated classification method and their morphological
appearance.  The parameters $n$ and $B$ are shown for all galaxies
with visual morphologies in Figure \ref{morphcheck}. A simple linear
relation separates early- and late-type galaxies: $B=0.065(n+0.85)$
(Fig. \ref{morphcheck}, \textit{solid line}).  Note that these
coefficients are taken from \citet{blakeslee06}, and are not chosen to
optimize the correspondence between the visual and $B-n$
classifications for these particular samples.  The correspondence
between our results, in particular for the $z\sim 0.03$ sample, and
the results by Blakeslee et al. illustrates the power of the $B-n$
method.  Once PSF smearing effects and noise properties are properly
taken into account, data sets with roughly similar spatial resolution
provide a unique measurement of galaxy morphology through the $B-n$
classification.  This extends the conclusions from Blakeslee et
al. who could not claim universality of their selection criteria
because of their homogeneous data sets and the small redshift range
($z=0.83-0.84$) of their samples.

The visual and $B-n$ classifications agree for 93\% of the galaxies in
the $z\sim 0.8$ sample (193 of 207) and for 89\% of the galaxies in
the $z\sim 0.03$ sample (179 of 200; see also Fig. \ref{morphcheck}).
Half of the mis-classifications have $n$- and $B$-values that put them
close to the criterion that separates the early types from the late
types (Fig. \ref{morphcheck}, \textit{solid line}). The disagreement
can be explained by the limited signal-to-noise ratio ($S/N$) of the
images, which shows that the true random uncertainty in the visual and
$B-n$ methods is $\sim$4\%. The other half of the disagreements are
mostly due to strong, central deviations from a S\'ersic profile
(probably point sources) or, for several galaxies in the $z\sim 0.03$
sample, large-scale deviations from a smooth profile that our fitting
method does not place within two effective radii.  Despite the
nonnegligible numbers of erroneous classifications ($\sim$10\%), the
net difference between the two classification methods in the ratio of
the numbers of early-type and late-type galaxies is less than 0.5\%.
This is true for both the $z\sim 0.03$ sample and the $z\sim 0.8$
sample.  Hence, there is virtually no systematic difference between
the visual and $B-n$ classifications for the samples as a whole.

\begin{figure}
\epsscale{1.2} 
\plotone{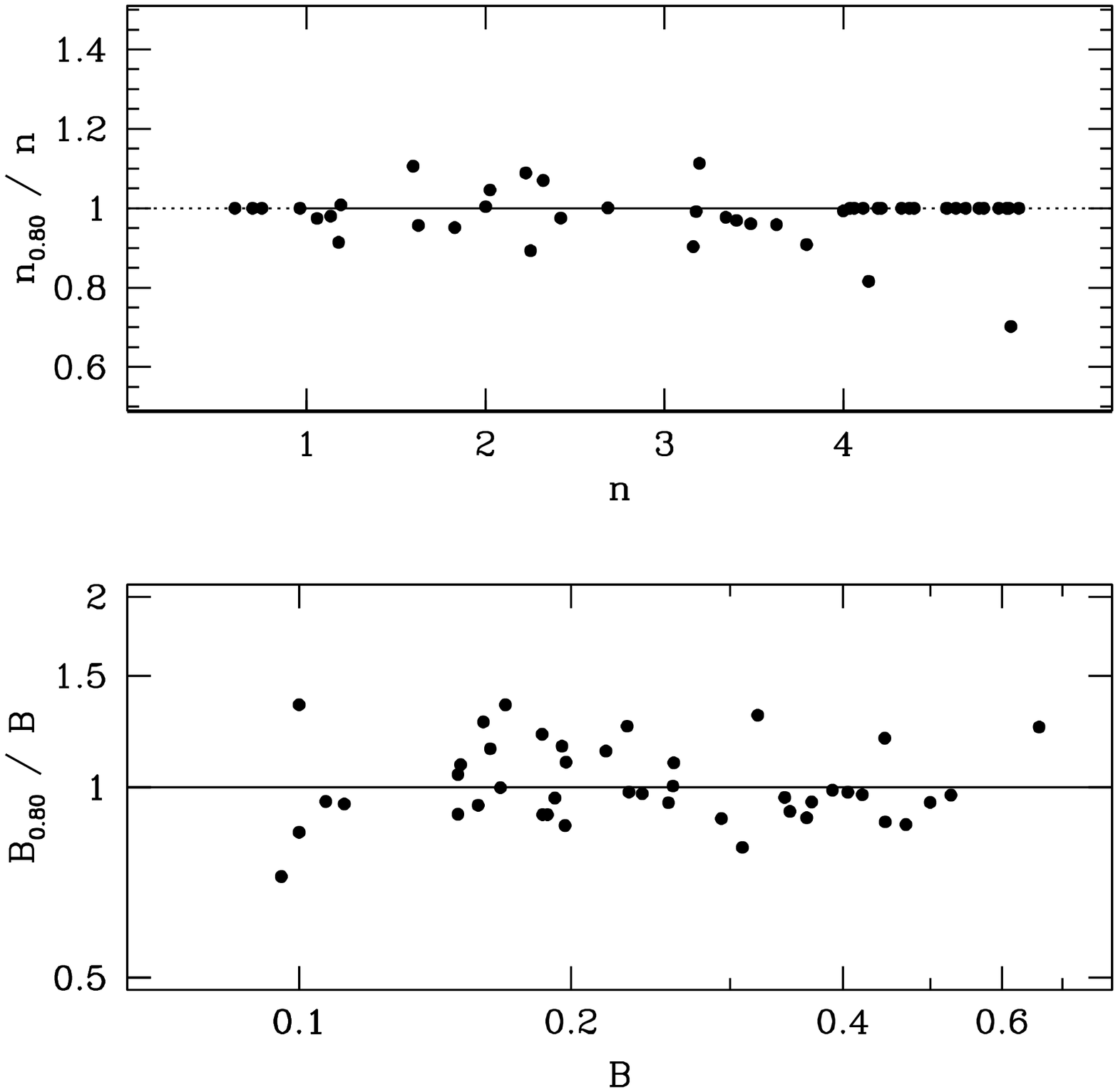} 
\caption{ S\'ersic parameter $n$ and bumpiness parameter $B$ for 50
 galaxies at redshifts $0.025<z<0.030$ as determined from the original
 SDSS $g$-band images vs. the difference between $n$ and $B$ as
 determined from simulated $z=0.80~$ $z$-band images of the same
 galaxies and the original values. There are no systematic differences
 between the original low-$z$ and simulated high-$z$ morphological
 indicators, which indicates that the morphologies as determined for
 our $z\sim 0.03$ and $\sim 0.8$ samples are not systematically
 different.}
\label{morphz_sdss}
\end{figure}

Even though the $z\sim 0.03$ and 0.8 samples have internally
consistent morphological classifications, there may be a systematic
difference between the two samples, which are located at very
different cosmological distances. Our morphological classifications
could depend on redshift, mainly because of the lower $S/N$ of the
high-$z$ galaxy images.  Because the scales of the PSF and pixels of
the SDSS and ACS data sets are very similar in terms of physical size
at the sample redshifts, we can test the redshift dependence of our
morphological classification.  We take 50 galaxies in our sample
spanning the full range in magnitude and with redshifts
$0.025<z<0.030$, such that the SDSS pixel scale corresponds to the
pixel scale of an ACS image of a galaxy at $z=0.80$.  We add noise to
the $g$-band SDSS images, taking into account $K$-corrections and the
relative depths of the SDSS and ACS images, and reapply our two
classification methods, which gives $n_{0.80}$ and $B_{0.80}$, the
values for $n$ and $B$ as inferred from the simulated images.  In
Figure \ref{morphz_sdss} we compare those with the original values of
$n$ and $B$, and we find that neither quantity shows a systematic
difference.  In addition, the visual morphologies of the simulated
images are not different from the original visual morphologies: all
features that decide on the morphology of the $z\sim 0.03$ galaxies
are still visible in the simulated $z=0.80$ images.  We conclude that
our classification methods do not suffer from systematic effects that
introduce significant differences between the morphologies of the
$z\sim 0.03$ and $z\sim 0.8$ galaxies.

\subsection{Local Density Estimates}\label{secdens}

Traditionally, local projected surface densities are computed by
measuring the distance to the $n$-th nearest neighbor brighter than,
for example, $M_{\rm{V}}+0.8z<-19.78$ \citep[e.g.,][]{postman05}.
However, since we work with mass-selected samples in this paper, it is
more consistent to measure the distance to the $n$th nearest neighbor
that is more massive than, in this case, $M>4\times
10^{10}~M_{\odot}$.  Below we discuss the difference between the two
approaches.

We derive the local galaxy densities for the 142 galaxies in our
$z\sim 0.8$ sample with spectroscopic redshifts.  We compute the
distance to the $n$-th nearest neighbor with a difference in
rest-frame radial velocity of $\Delta v<1000~\rm{km~s}^{-1}$ and more
massive than $M=4\times 10^{10}~M_{\odot}$ (or, alternatively, more
luminous than $M_{\rm{V}}+0.8z<-19.78$).  Here $n$ is the number of
galaxies with such properties in the CDF-S, with a maximum of 7.  The
reason for allowing $n$ to be smaller than 7 is the limited number of
galaxies at a given, spectroscopic redshift.  We note that for
galaxies with fewer than 7 neighbors (12 of 142 for the
luminosity-limited sample, 36 of 142 for the mass-limited sample) the
density estimate does not systematically change for any value $3<n<7$.
The inferred surface densities are corrected for edge effects and for
incompleteness of the spectroscopic redshift catalog, as inferred from
the spectroscopic + photometric redshift catalog.

\begin{figure}
\epsscale{1.2}
\plotone{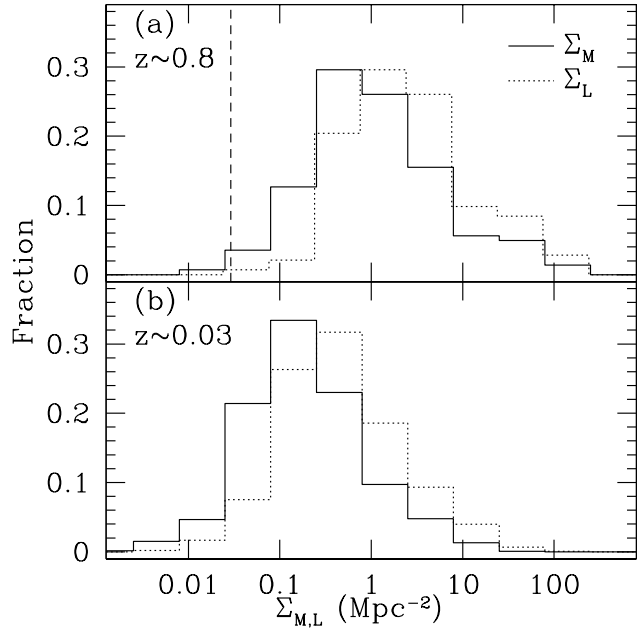}
\caption{ Distribution of local surface densities of the galaxies in
  our samples. (\textit{a}) Distribution of the $0.6<z<1.0$ sample,
  where the vertical dashed line indicates the value of $\Sigma$
  corresponding to one galaxy within the area covered by the survey
  data: objects with values of $\Sigma$ close to this line should be
  regarded as upper limits. (\textit{b}) Same as (\textit{a}), but for
  the $0.015<z<0.045$ sample. The solid histograms show the
  distributions of $\Sigma_{\rm{M}}$, the surface density as
  determined by regarding neighbors that are more massive than
  $M=4\times 10^{10}~M_{\odot}$. The dotted histograms show the
  distributions of $\Sigma_{\rm{L}}$, the surface density as
  determined by regarding neighbors that are brighter than
  $M_{\rm{V}}+0.8z=-19.78$.}
\label{Dhist}
\end{figure}

The distributions of $\Sigma_{\rm{M}}$ and $\Sigma_{\rm{L}}$ are shown
in Figure \ref{Dhist}; $\Sigma_{\rm{M}}$ is typically half of
$\Sigma_{\rm{L}}$, which is only a small difference considering the
large spread (3 orders of magnitude) of densities in the samples.  For
galaxies with only few neighbors, the computed values are likely upper
limits due to small area of the CDF-S.  Obviously, for galaxies
without close neighbors, the computed densities are certainly upper
limits.  The densities as computed here are, in principle, not
directly comparable with the computations for galaxies in cluster
environments \citep{postman05} as those use a wider redshift range to
search for neighbors to compensate for the large velocity dispersion
of the cluster members. A systematic difference of up to a factor of 2
may be expected due to the difference in bin width.  However, since
the environments under consideration have a range in density of
several orders of magnitude this does not compromise our inferences.

In a method analogous to that used for the $z\sim 0.8$ sample, we also
compute local galaxy densities by finding the distance to the seventh
nearest neighbor with $M=4\times 10^{10}~M_{\odot}$ (or
$M_{\rm{V}}+0.8z<-19.78$) and $\Delta v<1000~\rm{km~s}^{-1}$ for each
galaxy in the $z\sim 0.03$ sample.  The spectroscopic survey is 100\%
complete at these magnitudes, therefore corrections are not needed.
The distribution of $\Sigma$ is shown in Figure \ref{Dhist}.  As is
the case for the $z\sim 0.8$ sample, $\Sigma_{\rm{M}}$ and
$\Sigma_{\rm{L}}$ typically differ by a factor of 2 for the $z\sim
0.03$ sample (see Table \ref{tab2}).

There is a difference of $\sim$0.7 dex between the median values of
$\Sigma$ for the $z\sim 0.03$ and 0.8 samples.  This is mainly because
we use the angular diameter distance $D_{\rm{A}}$ to calculate the
physical distances between the galaxies, and not the transverse
co-moving distance $D_{\rm{M}}=D_{\rm{A}}(1+z)$. This provides, to
first order, redshift-independent values of the local density for
bound systems such as clusters, but for expansion-dominated regions
the density will be redshift dependent.  Using $D_{\rm{M}}$ would
provide a redshift-independent measurement of the density and would
lower the $z\sim 0.8$ densities by 0.5 dex with respect to the $z\sim
0.03$ densities. This demonstrates that the galaxies in both samples
are located in similar (co-moving) environments, besides an average,
apparent over-density of a factor of $1.5-2$ in the CDF-S which has
been noted before \citep[e.g.,][]{bundy05}.  This difference is not of
great interest for this study since we consider a range in density
that spans several orders of magnitude (see \S~\ref{secmdr}). Since we
are interested in the properties of the galaxy population as a
function of physical density, and to facilitate the comparison with
work on cluster galaxies, we use $\Sigma$ as calculated with
$D_{\rm{A}}$ in the remainder of this paper.

\section{The Evolution of the Early-Type Galaxy Fraction and the Morphology-Density Relation}

In \S~\ref{secetf} we calculate the evolution of the early-type galaxy
fraction in our field samples. In \S~\ref{secmdr} we compare this with
samples of cluster galaxies over the same redshift range and determine
the evolution of the MDR.

\subsection{The Field Early-Type Galaxy Fractions at $z\sim 0.8$ and 0}\label{secetf}

Our $z\sim 0.03$ sample consists of 2003 galaxies, the $z\sim 0.8$
sample of 207 galaxies. These volume-limited samples are complete down
to a stellar mass limit of $M=4\times 10^{10}~M_{\odot}$, about half
of the typical mass of a local field galaxy, or $M_{\rm{B}}\sim -20.1$
for red galaxies in the $z\sim 0.8$ sample. This luminosity limit for
red galaxies is similar to the $z=0.8-1$ luminosity limits of the
COMBO-17 \citep{bell04b}, DEEP2 \citep{faber07} and NDWFS
\citep{brown07} data sets used for recent red-sequence studies.

The early-type galaxy fraction at $z\sim 0.8$, for galaxies that are
more massive than $M=4\times 10^{10}~M_{\odot}$, is
$f_{\rm{ET}}=0.48\pm 0.07$.  This is based on the quantitative
morphological classification method described above, but we find the
same if we use the visual classifications (see also
\S~\ref{secmorph}).  If only galaxies with spectroscopic redshifts are
included (142 of 207 have spectroscopic redshifts), we find
$f_{\rm{ET}}=0.52$. The reason that the spectroscopic sample has a
slightly, but not significantly, higher early-type galaxy fraction is
that several of the spectroscopic campaigns specifically targeted red
objects.

\input{tab2.tex}

The main contributor to the error of 0.07 is the uncertainty in the
morphological classifications (0.055).  The second main source of
uncertainty is Poisson noise due to the limited sample size (0.036).
Smaller contributions include errors in photometric redshifts, color
transformations, and $M/L$ estimation.  All these individual
contributions are added in quadrature to obtain 0.07.

The early-type fraction at $z\sim 0.03$, for galaxies that are more
massive than $M=4\times 10^{10}~M_{\odot}$, is $f_{\rm{ET}}=0.43\pm
0.03$.  Essentially, the only significant contributor to the error is
the uncertainty in the morphological classifications, which is
slightly smaller than for the $z\sim 0.8$ sample because of the higher
$S/N$ of the images.

The early-type galaxy fractions in the mass-selected $z\sim 0.03$ and
0.8 samples are not significantly different (see Table \ref{tab2} and
Figure \ref{z_etf}).  This implies that if only two morphological
classes are considered, then the morphological composition of the
galaxy population did not change significantly between $z\sim 0.8$ and
the present, at least above our mass limit of $M=4\times
10^{10}~M_{\odot}$. If anything, the early-type fraction is higher at
$z\sim 0.8$ than at $z\sim 0.03$.

\subsection{Comparison with Cluster Galaxies: The Morphology-Density Relation}\label{secmdr}

\begin{figure}[b]
\epsscale{1.2}
\plotone{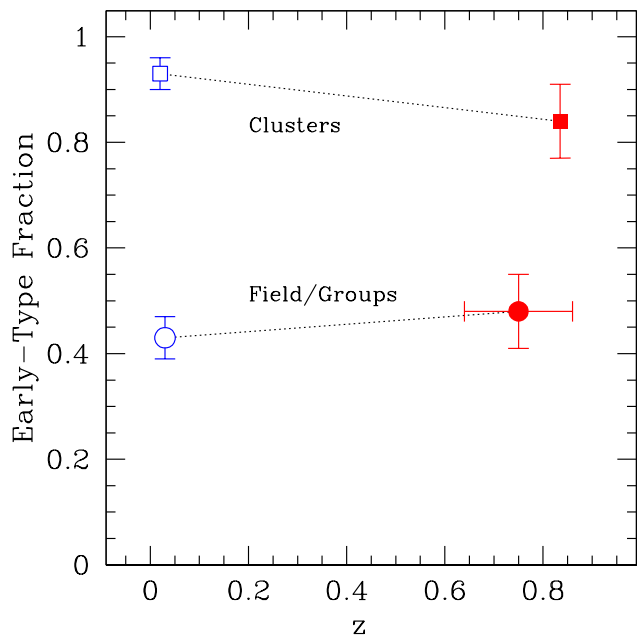}
\caption{ Early-type galaxy fraction as a function of redshift.  The
 circles indicate our samples of field galaxies ($\Sigma\lesssim
 50~\rm{Mpc}^{-2}$) at $z\sim 0.03$ (\textit{open blue circle}) and
 $z\sim 0.8$ (\textit{filled red circle}).  The squares indicate the
 samples of cluster galaxies ($\Sigma>50 ~\rm{Mpc}^{-2}$) from Paper I
 at $z=0.02$ (\textit{open blue square}, representing the Coma
 Cluster) and $z=0.83$ (\textit{filled red square}, representing the
 CL 0152 and MS 1054 clusters).  The fraction of early-type galaxies
 has not changed significantly since $z\sim 0.8$ for galaxy
 populations in similar environments.}
\label{z_etf}
\end{figure}

As we described above, the early-type galaxy fraction in our field
samples is $40\%-50\%$ at redshifts $0<z<1$.  In Figure \ref{z_etf} we
compare this with the results from Paper I, where we measure the
evolution of the cluster early-type fraction for galaxies down to the
same mass limit ($M=4\times 10^{10}~M_{\odot}$) and over the same
redshift range.  For galaxies more massive than $M=4\times
10^{10}~M_{\odot}$ the early-type galaxy fraction is constant with
redshift, both in the field and in massive clusters (see
Fig. \ref{z_etf}). The same MDR exists at $z=0$ and 0.8, with
early-type fractions of $40\%-50\%$ in low-density environments and
$>$80\% in high-density environments.

\begin{figure*}
\epsscale{1.1}
\plottwo{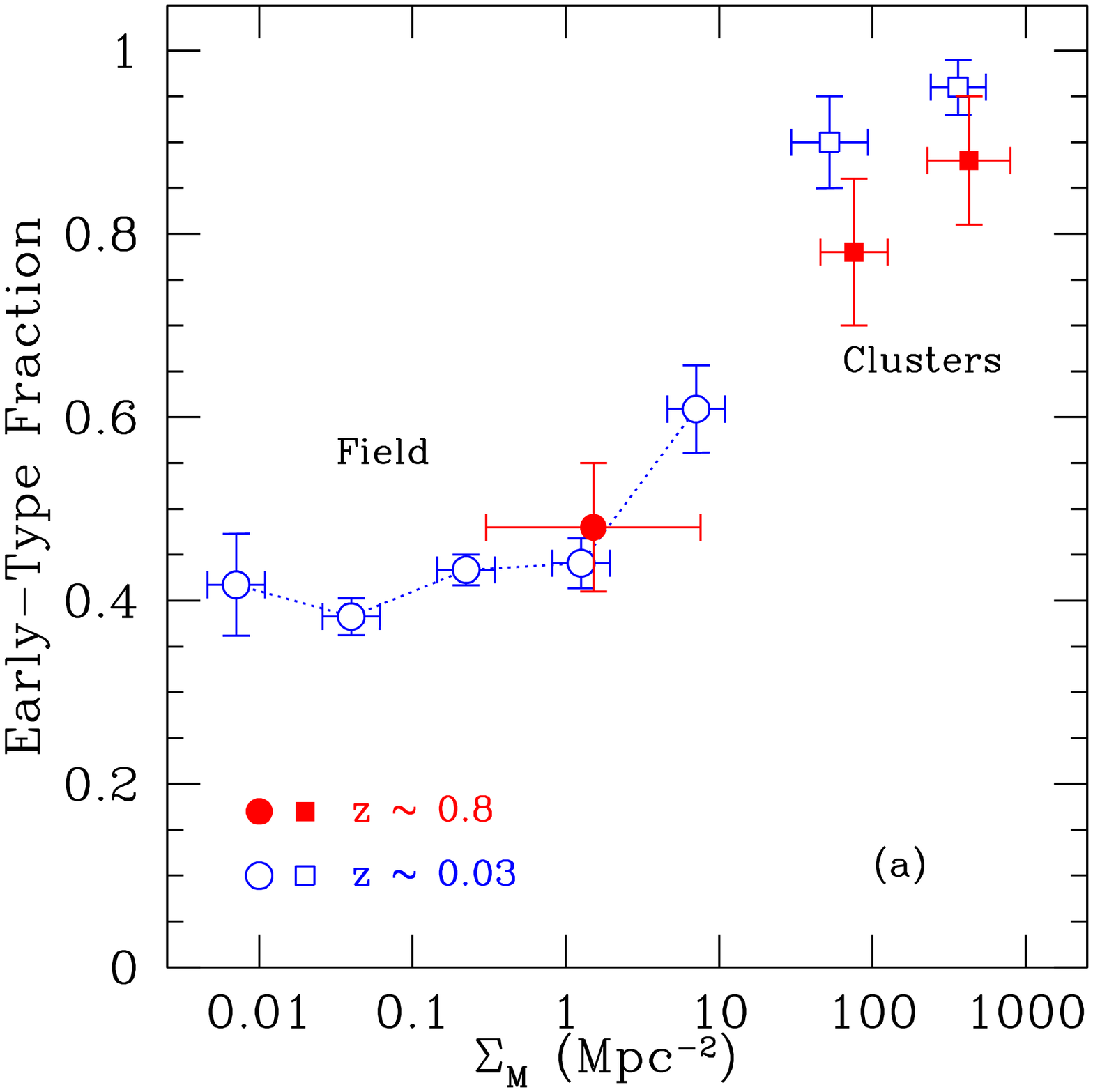}{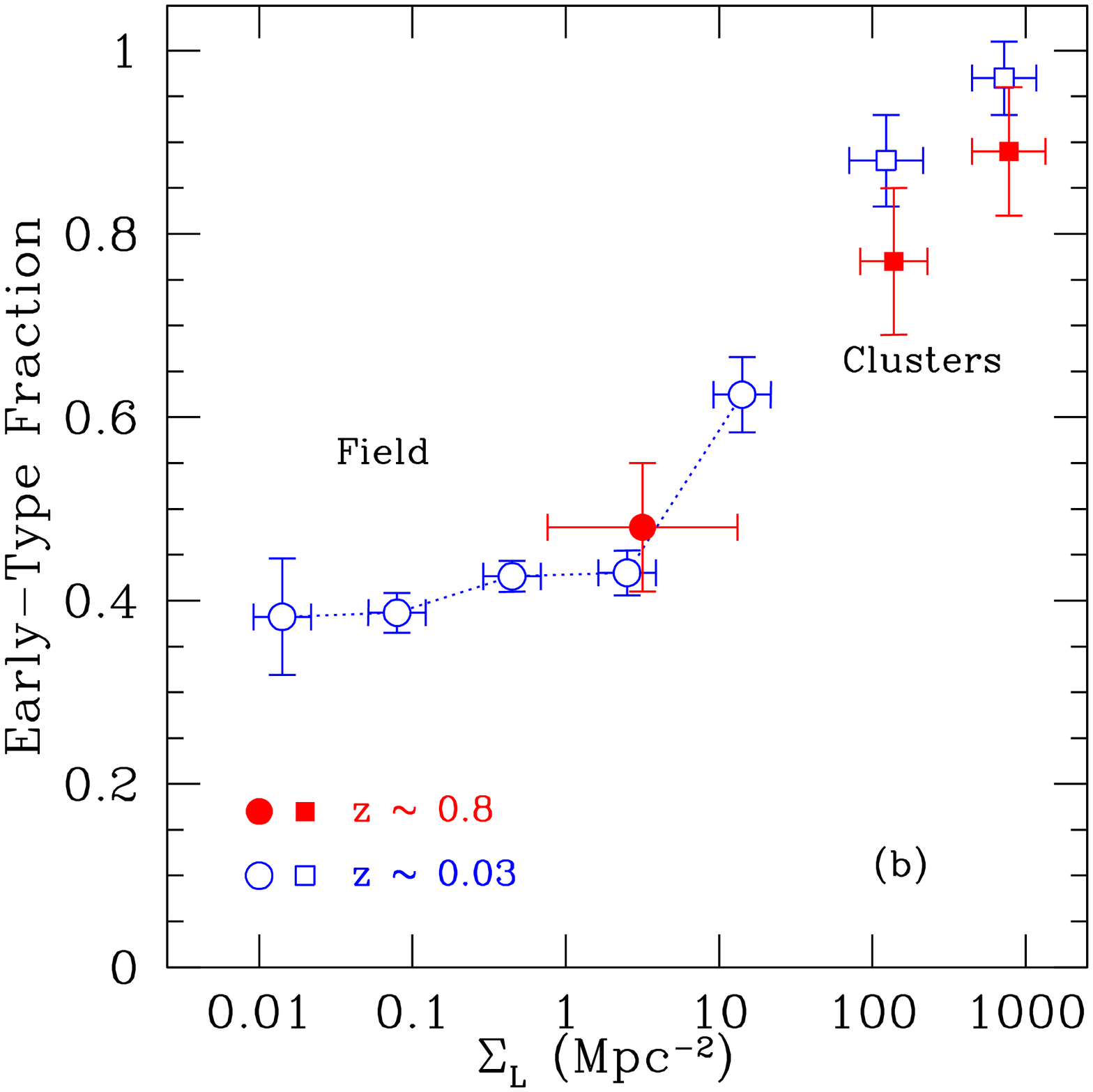}
\caption{ Morphology-density relation for mass-selected galaxies
 ($M>4\times 10^{10}~M_{\odot}$) at $z\sim 0$ and $z\sim 0.8$. The
 symbols are the same as in Figure \ref{z_etf}.  Our $z\sim 0.03$
 field sample (\textit{open blue circles connected by the dotted
 line}) have been split into five density bins.  Over almost 3 orders
 of magnitude in density the morphology-density relation has been in
 place since at least $z=0.8$ and has not evolved within the
 measurement errors.  The difference between the two panels is the use
 of $\Sigma_{\rm{M}}$ (the surface density of galaxies more massive
 than $M=4\times 10^{10}~M_{\odot}$, or $\sim0.5M_*$) in (\textit{a}),
 and the use of $\Sigma_{\rm{L}}$ (the surface density of galaxies
 more luminous than $M_{\rm{V}}+0.8z<-19.78$, or $\sim0.5L_*$) in
 (\textit{b}). Other than a shift of 0.3 dex in density the use of
 $\Sigma_{\rm{M}}$ and $\Sigma_{\rm{L}}$ is interchangeable.}
\label{morphdens}
\end{figure*}

The cluster samples from Paper I consist of galaxies with local
densities $> 50~\rm{Mpc}^{-2}$, our field samples consist of galaxies
with densities $< 50~\rm{Mpc}^{-2}$. In Figure \ref{morphdens} we show
the MDR at redshifts $z\sim 0.03$ and $\sim0.8$.  In this figure we
split our $z\sim 0.03$ field sample of 2003 galaxies into five density
bins, ranging from 0.01 to 10 Mpc$^{-2}$.  Our $z\sim 0.8$ field
sample is represented by a single data point, because of the limited
sample size. The average logarithm of the density of the 142 galaxies
with spectroscopic redshifts is used to estimate the density for the
sample as a whole.  The MDR has not evolved significantly since $z\sim
0.8$ for galaxies more massive than $M=4\times 10^{10}~M_{\odot}$. The
early-type galaxy fraction in regions with a given local density has
remained constant over the past $\sim$7 Gyr.

At first sight, this result seems to be at odds with previous studies
that reported significant evolution of the early-type galaxy fraction
\citep{smith05,postman05}. However, those analyses are based on
luminosity-selected samples, and, as is shown in Paper I, the observed
evolution is driven by blue, star-forming, low-mass galaxies. Samples
that are complete in terms of stellar mass do not show evidence for
evolution in the early-type fraction.  The evolution in the early-type
fraction seen in luminosity-selected samples is driven by the
evolution of the fraction of S0 galaxies, whereas the fraction of E
galaxies is roughly constant
\citep{dressler97,treu03,postman05,desai07}. It remains to be seen how
the relative fractions of E and S0 galaxies evolve in mass-selected
samples.

The MDR in the $z\sim 0.03$ data set shown in Figure \ref{morphdens}
suggests that the early-type fraction is low ($\sim$40\%) for all
local surface densities $\Sigma_{\rm{M}}\lesssim 1~\rm{Mpc}^{-2}$.
The transition to very high early-type fractions ($\sim$100\%) happens
at $1~\rm{Mpc}^{-2}<\Sigma_{\rm{M}}<100~\rm{Mpc}^{-2}$.  This suggests
that there is a critical density above which galaxies are
progressively more likely to undergo morphological transformations, as
was earlier noted by \citet{goto03}.  Despite the small area and the
small dynamic range in densities, there is evidence for this
transition in our $z\sim 0.8$ sample: if we take the galaxies with
spectroscopic redshifts in the highest 20th percentile of the density
distribution (typically, $\Sigma_{\rm{M}}\sim 30~\rm{Mpc}^{-1}$), we
find an early-type fraction of $f_{\rm{ET}}=0.8\pm 0.1$; for galaxies
in the lowest 20th percentile (typically, $\Sigma_{\rm{M}}\sim
0.2~\rm{Mpc}^{-1}$), we find $f_{\rm{ET}}=0.4\pm 0.1$. Without
over-emphasizing these numbers because the sub-sample with
spectroscopic redshifts is not necessarily representative of the whole
$z\sim 0.8$ sample, we do note that these values follow the same trend
as seen in the $z\sim 0.03$ sample (see Fig. \ref{morphdens}).

\section{Discussion}

\subsection{The Evolution of the Star Formation Rate}\label{secsf}

Our conclusion that for galaxies with masses $M>4\times
10^{10}~M_{\odot}$ the early-type galaxy fraction has not
significantly changed in any environment over the past 7 Gyr has to be
reconciled with the well-constrained decrease of the star formation
rate (SFR) density by an order of magnitude since $z\sim 1$
\citep[see, e.g.,][]{madau96,wolf03,lefloch05,bell05,perez05,zheng06}.
It may seem counter-intuitive to have the SFR decrease rapidly without
a change in the morphological mix of the galaxy population. On the
other hand, a non-changing early-type galaxy fraction does not
necessarily imply that the galaxy population itself does not
evolve. Below we investigate the star formation activity in our
samples.

We use the very deep, publicly available GOODS\footnote{The Great
Observatories Origins Survey (GOODS) data are available at
http://data.spitzer.caltech.edu/popular/goods/Documents/goods\_dr3.html}
$24~\mu\rm{m}$ imaging from the Multiband Imaging Photometer for
\textit{Spitzer} \citep[MIPS;][]{rieke04} to verify the star formation
activity in our $z\sim 0.8$ sample of mass-selected galaxies. Total
$24~\mu\rm{m}$ fluxes are measured by fitting the MIPS PSF to the
positions of objects detected in the $3.6~\mu\rm{m}$ band of the
Infrared Array Camera \citep[IRAC;][]{fazio04} on board
\textit{Spitzer}.  This technique reduces the effect of blending and
allows for more accurate photometric measurements \citep[for more
details, see][]{wuyts07}.  The $24~\mu\rm{m}$ flux is used to estimate
the bolometric infrared luminosity $L_{\rm{IR}}$ via the conversions
from \citet{chary01}. Subsequently, we use $L_{\rm{IR}}$ to estimate
the SFR \citep{kennicutt98}, where the SFR is scaled down by 0.15 dex
to account for the difference between the standard Salpeter IMF and
the diet Salpeter IMF that we use. For more details concerning the
$K$-correction and uncertainties (roughly a factor of 2 in the SFR) we
refer to \citet{vanderwel07}.  We conservatively use $50~\mu{\rm{Jy}}$
as the flux limit, as at lower flux levels the poor constraint on the
positions of $24~\mu\rm{m}$ sources requires verification of the
optical counterpart for each individual object, which is beyond the
scope of this paper.

\begin{figure*}
\epsscale{1.1}
\plottwo{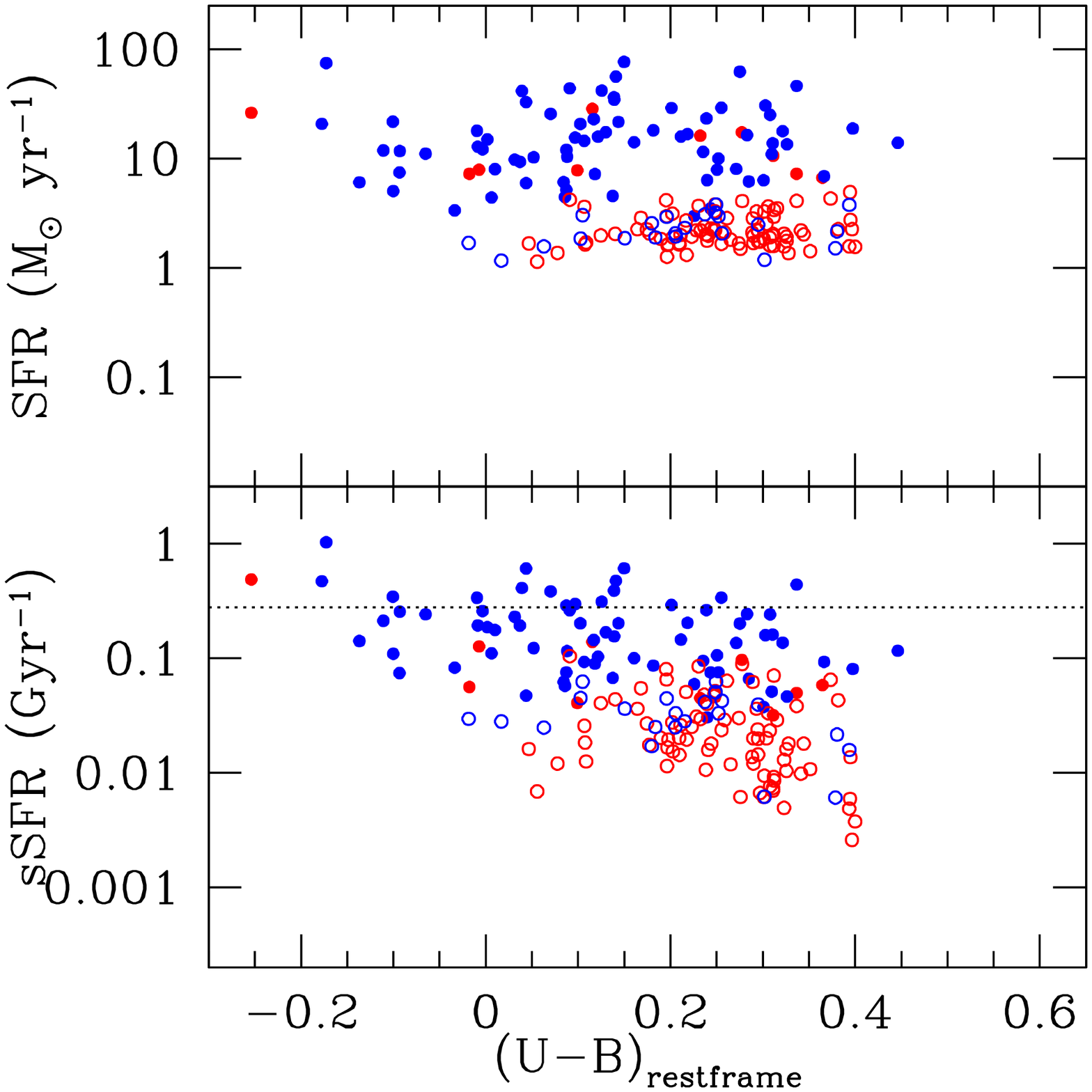}{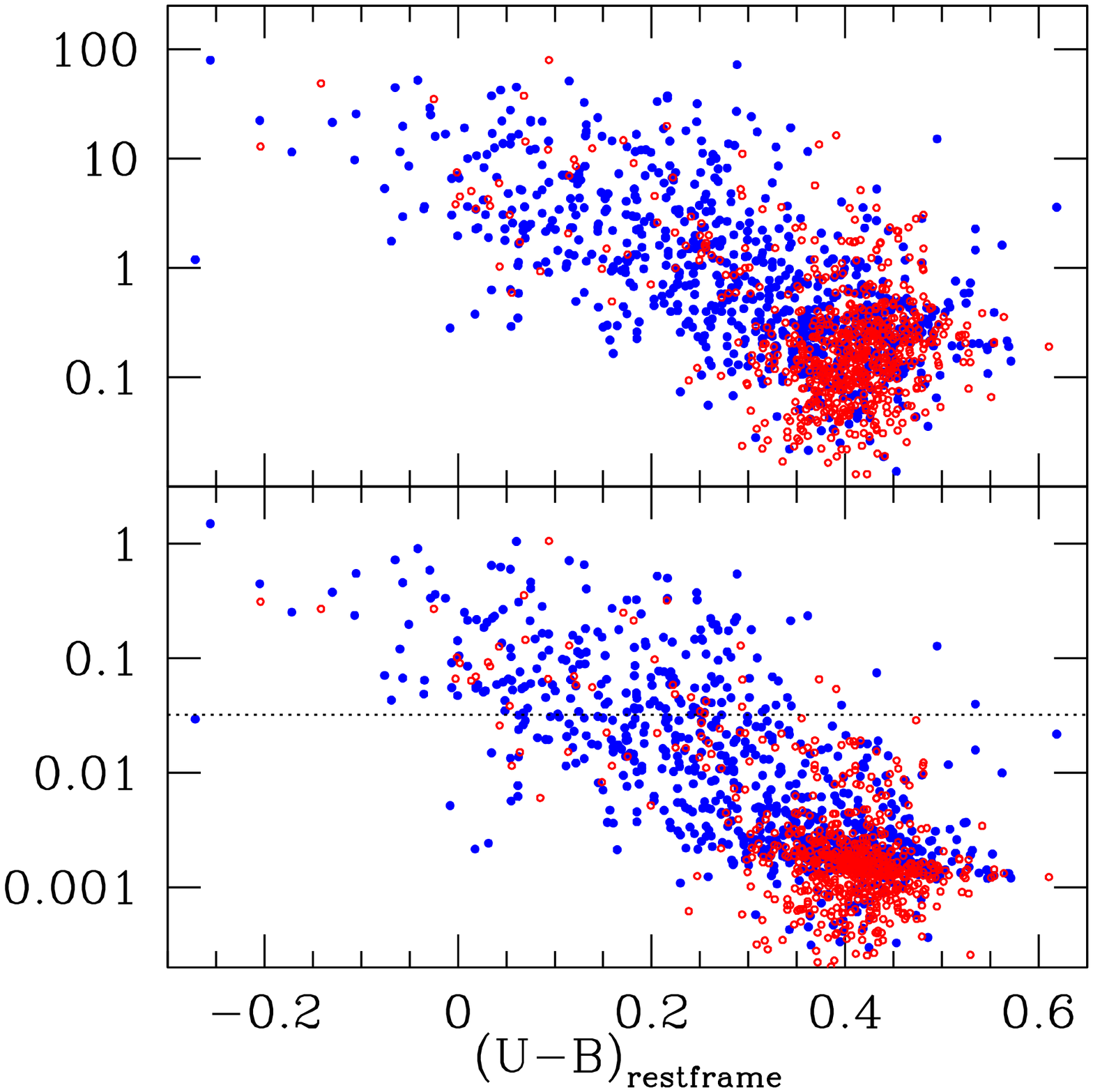}
\caption{ \textit{Top left:} SFR (derived from $24~\mu\rm{m}$
 photometry) as a function of rest-frame color for the $z\sim 0.8$
 sample.  The red circles are early-type galaxies, the blue circles
 are late-type galaxies. Filled circles are objects with significant
 ($>5\sigma$) $24~\mu\rm{m}$ detections, open circles are objects
 without significant detections (upper limits are shown).
 \textit{Bottom left:} For the same sample as in the top left panel,
 the specific SFR as a function of color.  \textit{Top right:} SFR
 (derived from modeling optical emission lines; Brinchmann et
 al. 2004) as a function of color for the $z\sim 0.03$ sample. Open
 red circles are early-type galaxies, and filled blue circles are
 late-type galaxies.  \textit{Bottom right:} For the same sample as in
 the top right panel, the specific SFR as a function of color. For
 both the low- and high-redshift samples, the star formation activity
 is low in early-type galaxies.  SFRs are higher for late types, with
 a marked decrease in SFR for red late-type galaxies between $z=0.8$
 and the present. The dotted lines in the two bottom panels indicate
 the expected specific SFRs associated with galaxies that will
 double/have doubled their stellar masses between $z=1$ and the
 present under the assumption that their SFRs decline (exponentially
 as a function of redshift) by an order of magnitude between $z=1$ and
 the present. The decreased star formation activity in our $z\sim
 0.03$ sample with respect to our $z\sim 0.8$ sample is consistent
 with the decline of the cosmic SFR.}
\label{UB_SFR}
\end{figure*}

In Figure \ref{UB_SFR} we show the SFR and the SFR per unit stellar
mass, the specific SFR, as a function of rest-frame $U-B$ color for
the 187 (of 207) $z\sim 0.8$ galaxies with MIPS coverage. Of the
early-type galaxies, 11\% are detected at $50~\mu{\rm{Jy}}$, at least
half of which are likely active galactic nuclei \citep{vanderwel07}.
Of the late-type galaxies, 76\% have fluxes $>50~\mu{\rm{Jy}}$, the
majority of which are most likely due to star formation. The typical
SFR of a late-type galaxy is $\sim 20~M_{\odot}~\rm{yr}^{-1}$, whereas
that of an early type is $\lesssim 5~M_{\odot}~\rm{yr}^{-1}$.  The
specific SFRs are $\lesssim 10^{-10}~\rm{yr}^{-1}$ for early types and
$\lesssim 10^{-9}~\rm{yr}^{-1}$ for late types.  These values imply an
increase in stellar mass of less than 10\% per Gyr for the early-type
galaxies and up to 100\% for the late types (see Fig. \ref{UB_SFR}).
These findings are consistent with the work by \citet{bell05}, who
show that the star formation activity at $z\sim 0.7$ is mainly due to
star formation in normal spiral galaxies with masses $>2\times
10^{10}~M_{\odot}$.

\citet{brinchmann04} derived SFRs for galaxies in the SDSS by modeling
their optical emission lines. The most robust determination of the SFR
for nearby galaxies such as those in our $z\sim 0.03$ sample is that
measured within the spectroscopic fiber of the SDSS spectrograph. We
use the stellar mass derived within the same fiber aperture
\citep{kauffmann03a},\footnote{The stellar masses and SFRs are
publicly available at http://www.mpa-garching.mpg.de/SDSS/DR4/} and
scale the SFR, comparing the fiber stellar mass with our total stellar
mass (\S~\ref{secsdss}). We show the SFR and specific SFR in Figure
\ref{UB_SFR}.  The global trend is similar to that observed at $z\sim
0.8$, with early-type galaxies showing lower star formation activity
than late-type galaxies.

It should be kept in mind that SFRs of the $z\sim 0.03$ galaxies are
determined from optical emission lines as measured with a $3''$ fiber,
such that extended star formation can be missed, and a systematic
difference between the total SFRs of the $z\sim 0.8$ galaxies is
created. In addition, it is well known that IR-derived and emission-
line-derived SFRs can be intrinsically different and depend on
metallicity and the amount of extinction.  Therefore, we have to
restrict ourselves to the description and interpretation of
evolutionary trends of at least an order of magnitude.

Despite the large uncertainties and the high upper limit of
$5~M_{\odot}~\rm{yr}^{-1}$ for the $z\sim 0.8$ sample, Figure
\ref{UB_SFR} shows that the overall star formation activity decreases
between $z\sim 0.8$ and the present (see also Table \ref{tab2}).  One
prominent difference between the $z\sim 0.03$ and $\sim 0.8$ samples
is the marked decrease of star formation activity by 2 orders of
magnitude in late-type galaxies that are located on the red sequence.
In the local universe, their SFRs are as low as those of early-type
galaxies, whereas at $z\sim 0.8$ their SFRs are at least an order of
magnitude higher than those of the early types.

In Figure \ref{UB_SFR} we indicate the specific SFR expected for
galaxies that double their stellar mass between $z=1$ and the present,
assuming an exponential decline in SFR by a factor of 10 over that
period. Such a level of star formation, both at $z\sim 0.03$ and at
$z\sim 0.8$, coincides with the typical specific SFR of late-type
galaxies.  The decrease in star formation activity as observed in our
samples follows the cosmic average trend of decline by an order of
magnitude since $z\sim 1$. At the same time, the morphological mix has
remained unchanged, at least, if only two morphological classes are
considered. These results are therefore not mutually exclusive and
have to be reconciled with each other.

\begin{figure*}
\epsscale{1.1} 
\plottwo{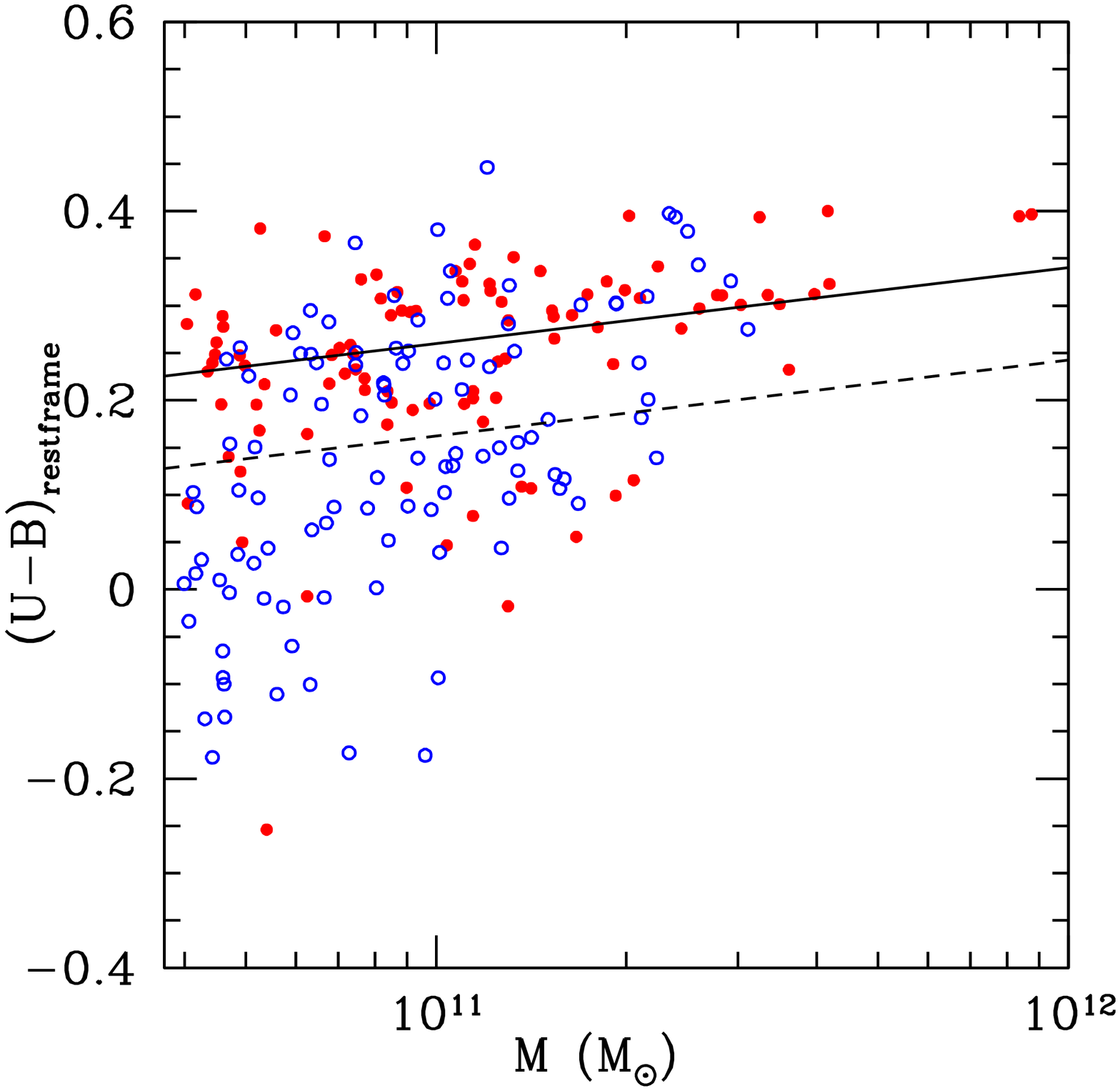}{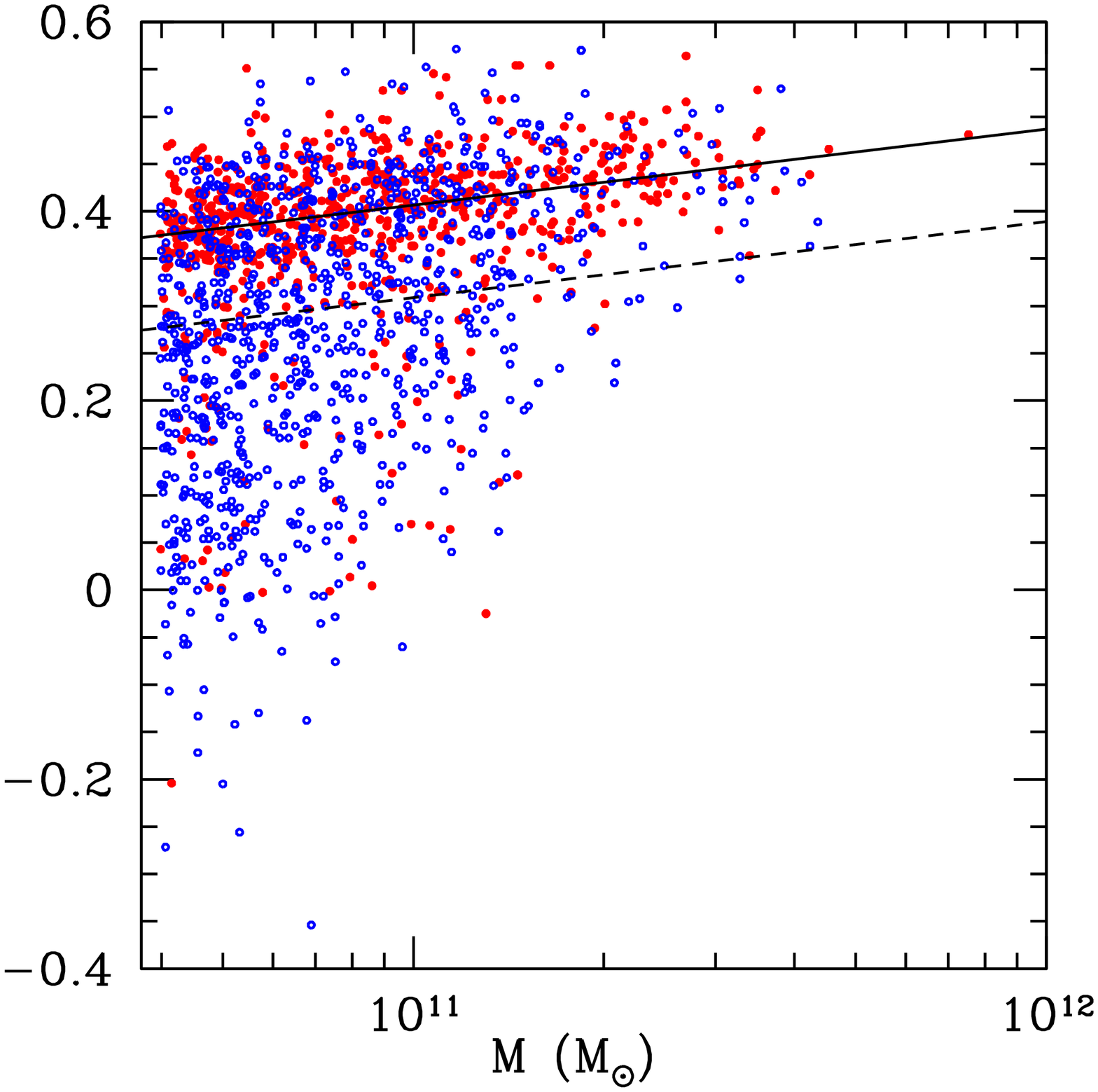}
\caption{ Stellar mass vs. rest-frame $U-B$ color for the $z\sim 0.8$
 sample (\textit{left}) the $z\sim 0.03$ sample (\textit{right}).
 Early-type galaxies are indicated by filled red circles symbols,
 late-type galaxies by open blue circles. Color evolution is readily
 visible, with the red sequence being significantly bluer at $z\sim
 0.8$ than at $z\sim 0.03$. The solid lines indicate the best fits to
 the red sequence as defined by the early-type galaxies, and the
 dashed lines separate blue and red galaxies (see text for an
 explanation). The relative numbers of red galaxies in the $z\sim
 0.03$ and $\sim 0.8$ samples are not significantly different (see
 also Table \ref{tab2}). In addition, the fraction of red galaxies
 with late-type morphologies is similar in both samples.}
\label{MCD}
\end{figure*}

\subsection{The Red Sequence and the Evolution of the Stellar Mass Density}\label{seccolormass}

In Figs. \ref{MCD} we show the rest-frame $U-B$ color distribution as
a function of stellar mass for the low- and high-redshift samples.
The $U-B$ colors of the $z\sim 0.8$ sample are corrected for evolution
within the sample redshift range $0.6<z<1.0$, assuming
$\Delta(U-B)=0.22(z-0.8)$ (see below).  This is a small effect of
typically 0.02 mag. We define the early-type galaxy red sequence for
the $z\sim 0.03$ sample by a linear fit, with the slope and the zero
point as free parameters, and iteratively rejecting $3\sigma$ outliers
(8\% are rejected, virtually all blueward of the red sequence). In
order to minimize systematic effects, we define the red sequence for
the $z\sim 0.8$ sample with the same slope (0.08 mag dex$^{-1}$) and
scatter (0.05 mag), with only the zero point as a free parameter.  The
red sequence for both samples is shown in Figure \ref{MCD}.  Color
evolution between $z\sim 0.8$ and $\sim 0.03$ of 0.15 mag, or 0.20 mag
per unit redshift, is readily visible. We also fit the slope and
scatter of the $z\sim 0.8$ red sequence independently, resulting in a
higher scatter (0.08 mag) and a steeper slope (0.11 mag dex$^{-1}$).

\begin{figure*}
\epsscale{1.1}
\plottwo{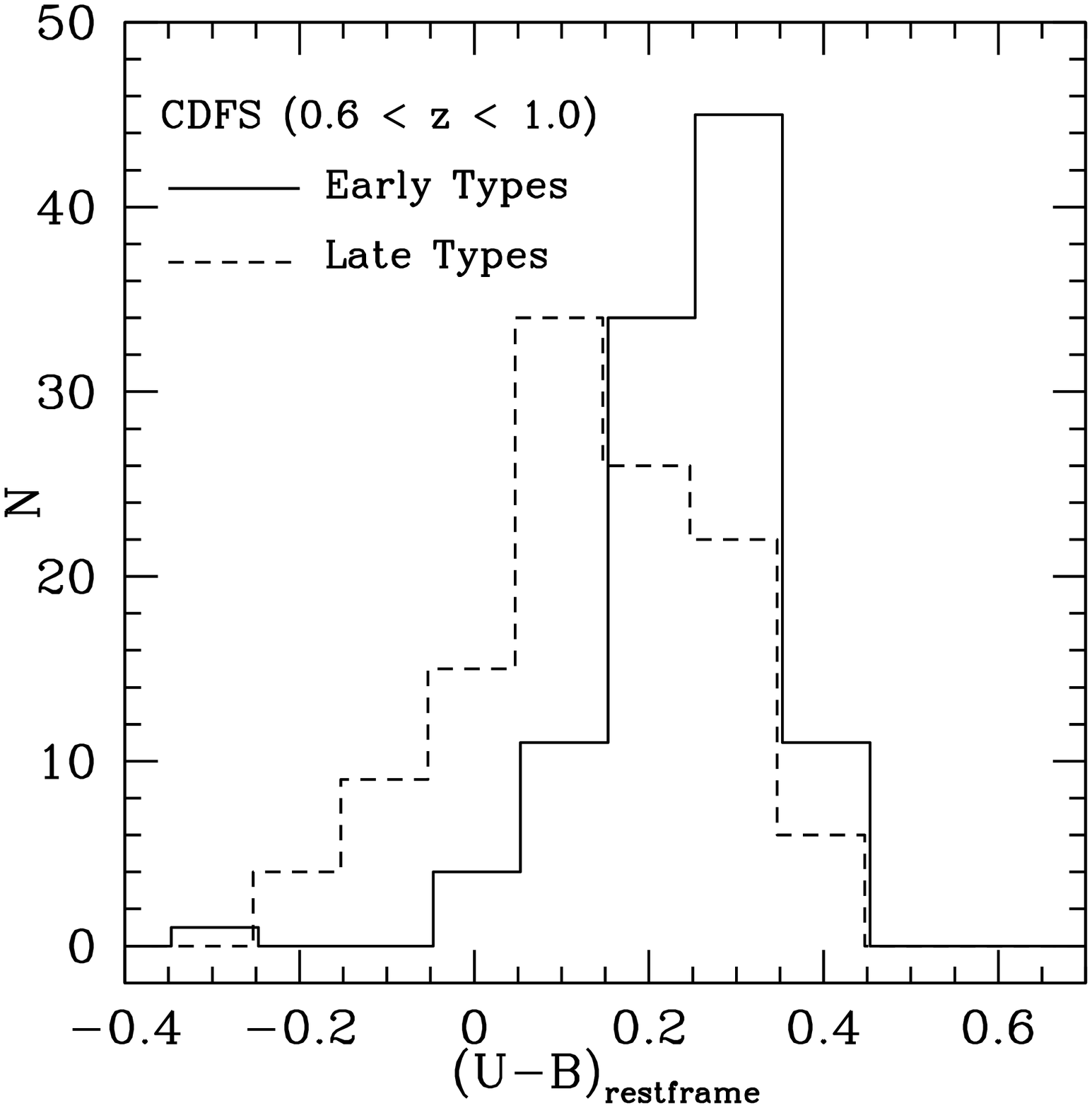}{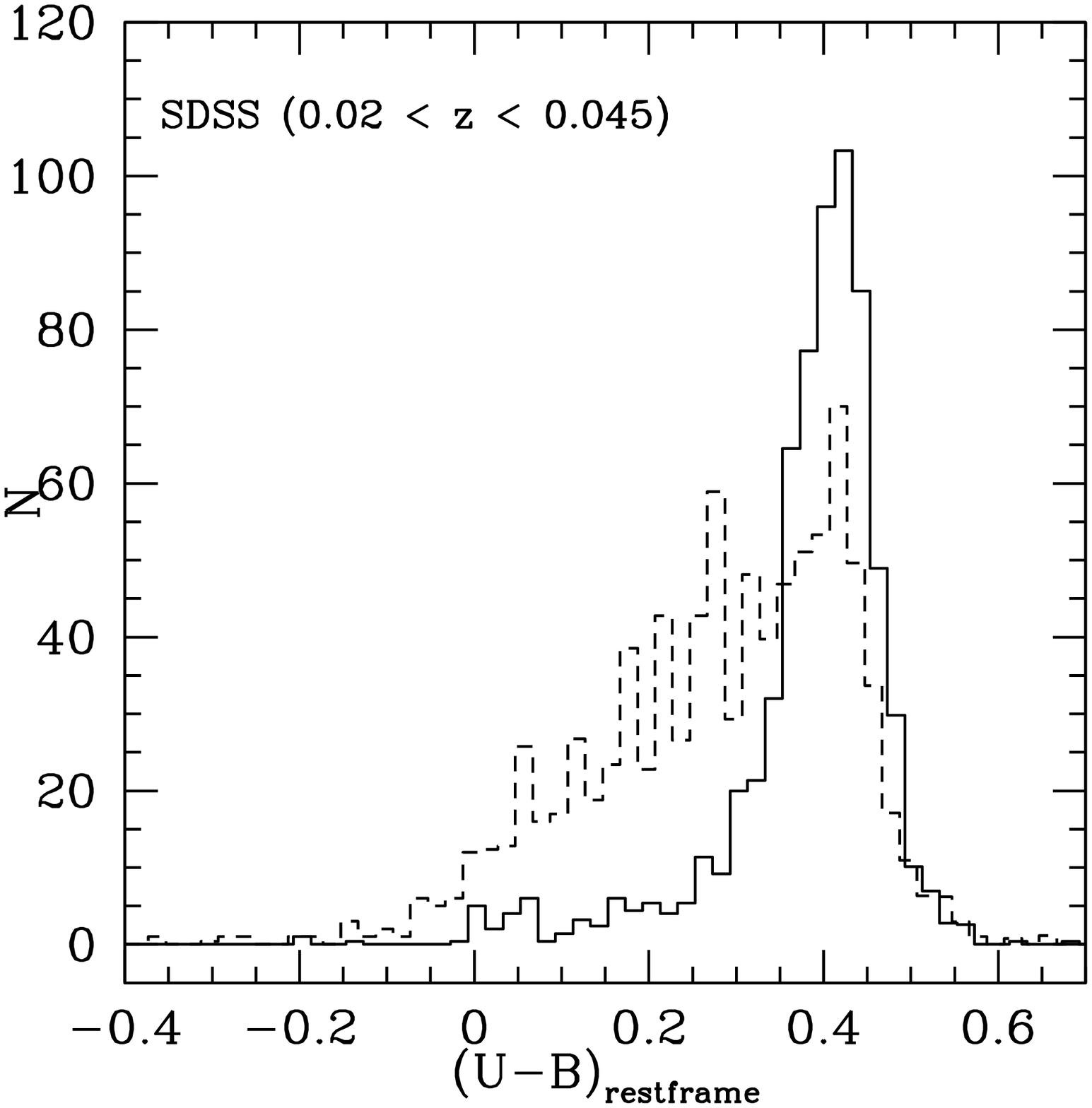}
\caption{ Rest-frame $U-B$ color distribution of the mass selected
  $z\sim 0.8$ (\textit{left}) and $z\sim 0.03$ (\textit{right})
  samples. The solid lines indicate the early-type galaxies, and the
  dashed lines indicate the late-type galaxies. The relative numbers
  of red galaxies and the relative numbers of red galaxies with
  late-type morphologies are not significantly different at $z\sim
  0.03 $ and $\sim 0.8$. }
\label{UB_N}
\end{figure*}

For both samples we define blue galaxies as galaxies with colors more
than 2 $\sigma$ below the red sequence (with $\sigma=0.05$ as defined
above; see Fig. \ref{MCD}). With this definition, we find similar
red-galaxy fractions of 64\% and 68\% for the $z\sim 0.8$ and $\sim
0.03$ samples, respectively (see Table \ref{tab2}).  Like the
early-type fraction, the red-galaxy fraction shows no strong evolution
with redshift. The red-galaxy fraction is significantly higher than
the early-type fraction in both samples \citep[see
also][]{bundy06}. This is due to the presence of many late-type
galaxies on the red sequence: $38\%\pm 8\%$ for the $z\sim 0.8$ sample
and $45\%\pm 3\%$ for the $z\sim 0.03$ sample of the red galaxies are
late types (see also Fig. \ref{UB_N}). Conversely, in either sample
less than 20\% of the blue galaxies have early-type morphologies.

Given the fact that the majority of the galaxies in both our samples
are located on the red sequence, our earlier conclusion that the
early-type galaxy fraction at a given density does not evolve with
redshift seems, at first sight, to be at odds with the observation
that the stellar mass density of red-sequence galaxies has doubled
since $z=1$ \citep[e.g.,][]{bell04b,brown07}.  The luminosity limit
for red galaxies at $z=1$ in our sample ($M_{\rm{B}}\sim -20$) is
similar as that of, e.g., \citet{bell04b} and \citet{brown07};
therefore, the increase in mass density found by those authors has to
take place above the mass threshold of our samples. In fact, even in
our samples, which cover only small volumes, we observe an increase in
stellar mass density from $z=0.8$ to the present.

If the mass function of galaxies evolves, it is not entirely
appropriate to compare galaxy samples at different redshifts down to
the same mass limit. If galaxies evolve by a factor of 2 in stellar
mass between $z=0.08$ and the present, the mass limit at $z=0.8$
should be chosen 0.3 dex lower than at $z=0$, otherwise many of the
progenitors in the $z=0$ sample will not be included in the $=0.8$
sample. We test this effect by choosing a mass limit for our $z\sim
0.03$ sample that is 0.3 dex higher ($M=8\times 10^{10}~M_{\odot}$)
than for the $z\sim 0.8$ sample. The early-type fraction increases
slightly, from $43\%\pm3\%$ to $48\%\pm3\%$. Due to the errors it is
not clear whether this agrees better with the $z\sim 0.8$ early-type
galaxy fraction ($48\%\pm7\%$), but it is striking that the values are
the same.

We conclude that the growth of the red-sequence galaxy population must
happen in such a way that the MDR is conserved. The accretion of red
galaxies onto the red sequence and/or the growth of galaxies on the
red sequence must preserve the relative number of late- and early-type
galaxies with masses $M>4\times 10^{10}~M_{\odot}$. The galaxy
population evolves along the MDR, whereas the MDR itself is a fixed
relationship between environment and morphological appearance.

Since many star-forming, late-type galaxies are located on the red
sequence, the stellar mass density of red-sequence galaxies will
increase through \textit{in situ} star formation.  The question is how
much this contributes to the observed increase in stellar mass density
of red galaxies.  If we make the assumption that the SFR of the
galaxies in our $z\sim 0.8$ sample decreases exponentially at the same
rate as the cosmic average, i.e., by an order of magnitude between
$z=1$ and the present, the average late-type galaxy on the red
sequence will increase its stellar mass by $20\%-80\%$. The large
uncertainty is caused by the absolute uncertainty in the IR-derived
SFR of a factor of 2 (see \S~\ref{secsf}).  Given such an increase in
stellar mass, there will be galaxies that are below our mass cutoff at
$z\sim 0.8$ but would not be during the present epoch. We can estimate
this effect by calculating the total stellar mass in red, late-type
galaxies in our $z\sim 0.03$ sample with masses less than $1.2-1.8$
times the threshold. Their contribution to the total stellar mass in
red galaxies is $2\%-11\%$, such that the total increase in the
stellar mass density through \textit{in situ} star formation in red
late types is a factor of $1.2-2$. This is close the observed
evolution of a factor of $\sim$2. One caveat is the possible evolution
with redshift of the color-$M/L$ relation, which we use to estimate
the stellar masses (see \S~\ref{secdatacdfs}). This may cancel out the
growth in stellar mass as estimated above.

With the hypothesis that merging does not play a role in shaping the
red sequence galaxy population, the above described process of star
formation in late types, but not in early types, will change the
morphological mix of a mass-selected sample of red galaxies. The
early-type fraction will then decrease with time or increase with
redshift. As mentioned in \S~\ref{seccolormass}, the fraction of
early-type galaxies on the red sequence is indeed consistent with an
increase from $z\sim 0.03$ to $\sim 0.8$ (from 55\% to 62\%). Even
though this difference is not significant, at least it is consistent
with a growth in stellar mass of late-type galaxies by a factor of
1.5.

These speculations depend on the assumption that a large reservoir of
lower mass, red galaxies is available to provide the necessary growth
of stellar mass through star formation on the red sequence. Given the
evidence for the opposite, i.e., the lack of faint, red galaxies at
significant look-back times \citep[e.g.,][]{tanaka05,delucia07},
truncation of star formation in blue galaxies is possibly required to
explain the growth of the red sequence. The above estimates of growth
through \textit{in situ} star formation merely serve as a cautionary
statement to indicate that a factor of 2 evolution in the red-galaxy
mass density is rather modest and can be accounted for in various
ways.

Of course, if, for some reason, current measurements overestimate the
evolution in the red-galaxy population, our results can be regarded as
evidence for passive evolution of red galaxies, in terms of both star
formation and mass assembly. We stress, however, that this is most
likely not the case, given the mounting evidence for considerable
evolution in the stellar mass density of red galaxies since $z=1$.

\subsection{Comparison with Other Studies}
Numerous studies have analyzed the morphological and structural
properties of galaxies in the local universe and at high
redshifts. Even though it is not feasible to discuss all previous
work, we put our work on the local galaxy population in the context of
several other studies and compare our results with those from several
recent efforts with the similar objective to measure the evolution
with redshift of the early-type fraction outside massive clusters.

\citet{goto03} derive the MDR for luminosity-selected galaxies at
$0.05<z<0.10$. Their morphological classifications are simply based on
the concentration of the light profile, and they find early-type
galaxy fractions of $\sim40\%-60\%$ at low densities, depending on the
strictness of the classification criterion for early types. This is
consistent with our results. \citet{goto03} note that at densities
below $\Sigma\sim 1~\rm{Mpc}^{-2}$ the early-type fraction changes
less rapidly with density than at higher densities. This trend is even
more pronounced in our mass-selected sample: at densities
$\Sigma<1~\rm{Mpc}^{-2}$ the early-type fraction is constant at $\sim
40\%$ and only increases at higher densities (see Fig.
\ref{morphdens} and \S~\ref{secmdr}).

The contributions of galaxies with different morphologies to the total
stellar mass density have been measured by \citet{bell03}, who find
that nowadays 76\% of the stars in galaxies more massive than
$M=4\times 10^{10}~M_{\odot}$ reside in early-type galaxies. In our
$z\sim 0.03$ sample, we find a much lower value of 47\%. The cause of
this difference is that Bell et al. use a single parameter (the
concentration index) as a proxy for morphology, which corresponds with
visual classifications 70\% of the time \citep[see
also][]{strateva01}. Our method uses the bumpiness parameter $B$ in
addition to the PSF-corrected concentration of the light profile,
quantified by the S\'ersic parameter $n$.  Thus, early-type galaxies
are required to be not only highly concentrated but also smooth.  This
extension increases the agreement with visual classifications to
$\sim$90\% (see \S~\ref{secmorph}).  This improvement is not due to
smaller random errors: the contribution to the total stellar mass
density of highly concentrated galaxies in our $z\sim 0.03$ sample
(those with $n>2.5$) is 70\%, in good agreement with the result by
\citet{bell03}, and higher than that of early-type galaxies as defined
by both visual classifications and by our $B-n$ method. Many of the
galaxies with $n>2.5$ have late-type morphologies, which is only
revealed through visual classifications or the use of an additional
parameter such as bumpiness.  As was demonstrated by
\citet{kauffmann04}, the concentration distribution of the galaxy
population provides crucial insight into the formation and evolution
of galaxies. However, this structural property should be distinguished
from morphology. The latter is related to the smoothness of the light
profile in addition to its concentration.  It may well be that
structure and morphology are correlated but distinct physical
parameters that are affected by, e.g., galaxy mass and environment in
different ways.

Recently, several studies have addressed the evolution with redshift
of the morphological mix of field galaxies.  \citet{abraham07} find
that for galaxies more massive than $M=4\times 10^{10}~M_{\odot}$ the
fraction of the stellar mass density residing in early-type galaxies
is $\sim$80\% at $z\sim 1$, similar to the numbers from \citet{bell03}
and our $n$-selected sample at low redshift.  Abraham et al.  use two
parameters to distinguish early types from late types: in addition to
the Gini coefficient (which, for early types and spiral galaxies,
closely corresponds to the concentration index) they use asymmetry as
a morphological indicator.  However, the addition of asymmetry does
not substantially improve the agreement with visual classifications
for galaxies with type Sb or later \citep[see, e.g.,][]{conselice03}.
The method used by \citet{abraham07} therefore effectively classifies
galaxies by concentration.  In our $z\sim 0.8$ sample, highly
concentrated galaxies (with $n>2.5$) contribute 72\% to the total
stellar mass density, similar to what Abraham et al. find
($\sim$80\%).  The stellar mass in early-type galaxies as defined by
our $B-n$ method or visual classifications is lower (57\%), consistent
with the results by \citet{bundy05}, which are also based on visual
classifications.  We conclude that the difference between our numbers
and those from \citet{abraham07} are due to the difference between the
classifying methods.  Note that the relative early-type galaxy mass
density does not evolve between $z\sim 1$ and the present, independent
of the applied morphological classification method.

\citet{capak07}, who use the Gini coefficient to determine
morphologies, find a significant decrease with redshift (in the range
$0.4<z<1.2$) of the early-type fraction in high-density regions (with
$\Sigma>100~\rm{Mpc^{-2}}$).  At lower densities (with
$\Sigma<50~\rm{Mpc^{-2}}$) they find a constant early-type fraction
between $z=0.4$ and 1.2 that is consistent with the early-type
fraction that we find.  However, when they compare their results with
the local early-type fraction, measured with different classification
methods, they do find significant evolution, also at the lowest
densities. The sample constructed by \citet{capak07} is selected by
luminosity, and is therefore not directly comparable with our
sample. As was demonstrated above, selecting early-type galaxies by
Gini coefficient or concentration index alone overestimates the number
of early-type galaxies as compared to visual classifications or our
$B-n$ method. This, combined with the difference between luminosity-
and mass-selected samples, may conspire to yield an early-type
fraction similar to that we find for visually classified galaxies in a
mass-selected sample.

We conclude that our results are either consistent with previous work
or can be explained by the differences in approach. We note that the
GOODS/ACS imaging that we use to determine galaxy morphologies at
$z\sim 0.8$ is much deeper than then imaging used by \citet{capak07}
and effectively as deep as the data used by \citet{abraham07} in terms
of $S/N$ for red galaxies at $z\sim 1$. Moreover, our study is the
first of its kind to quantify galaxy morphologies in the local
universe and at higher redshifts in an internally consistent manner.

\section{SUMMARY}

In this paper we examine the early-type galaxy (E+S0) fraction and its
evolution for stellar-mass-selected, volume-limited samples of
galaxies at redshifts $0<z<1$.  At low redshift we estimate stellar
masses and determine morphologies for 2003 galaxies at redshifts
$0.02<z<0.045$ in the SDSS, complete down to a mass of $M=4\times
10^{10}~M_{\odot}$, 60\% of the mass of a typical ($L_*$) galaxy.  In
addition, we construct a similar sample of 207 galaxies in the CDF-S
at redshifts $0.6<z<1.0$, complete down to the same mass limit.  The
stellar mass estimates are based on the relation between color and
$M/L$, which has been proved to work without significant systematic
effects with respect to kinematic mass measurements at low and high
redshift.  Morphologies are determined with an automated method, based
on the S\'ersic parameter and the bumpiness of the residual.  We
compare this method with the traditional visual classifications and
find no systematic differences when we split the sample into two
morphological classes (Sp+Irr and E+S0).  We estimate the local
surface density with an $n$th nearest neighbor method. The density for
the galaxies in our samples is typically $\Sigma \sim 1~\rm{Mpc^{-2}}$
and ranges from $\sim$0.01 to $100~\rm{Mpc^{-2}}$. The sample
characteristics are summarized in Table \ref{tab2}.

We find that for galaxies with masses higher than $M=4\times
10^{10}~M_{\odot}$ the early-type fraction in the field and group
environment probed by our samples has not changed significantly
between $z=0.8$ and the present. The early-type fraction is
$43\%\pm3\%$ for the $z\sim 0.03$ sample, and $48\%\pm7\%$ for the
$z\sim 0.8$ sample. When we combine this with the unchanging
early-type fraction of $>80\%$ in dense ($>100~\rm{Mpc}^{-2}$) cluster
environments over the same redshift range and down to the same mass
limit (Paper I), we find that the MDR has not evolved significantly
between $z\sim 0.8$ and the present over at least 3 orders of
magnitude in density.

In both our $z\sim 0.03$ sample and our $z\sim 0.8$ sample,
$65\%-70\%$ of the galaxies (see also Table \ref{tab2}) are located on
the red sequence, in agreement with previous work. In addition,
$65\%-70\%$ of the galaxies in both samples have S\'ersic indices
$n>2.5$.  Our samples follow the well-constrained trends that the
stellar mass density of red galaxies has increased and the global star
formation activity has decreased between $z\sim 1$ and the present.
In fact, star formation in red galaxies could contribute significantly
to the growth in the stellar mass density of red galaxies without
strong evolution in the early-type galaxy fraction.

Future work will include a more detailed analysis of the morphological
mix, e.g., the separation of the late types into irregular and spiral
galaxies and the separation of the early types into E and S0 galaxies.
Furthermore, studies in larger fields with high-resolution imaging
will be used to measure the shape of the MDR at low densities in more
detail.  In particular, it is important to study intermediately dense
regions, i.e., the infalling regions around clusters. At those
densities the morphological mix of the galaxy population appears to
undergo the most rapid change as a function of density.  A crucial
step forward in understanding the role of the environment and the role
of the internal properties of galaxies in shaping their stellar
populations and morphologies will be to include the evolution of the
density itself in analyses such as those carried out in this paper.
Another challenge will be to establish at what redshift the MDR
emerges, whether this coincides with the emergence of the red
sequence, and how this relates to the buildup of the stellar mass
function.

\acknowledgements { We thank the referee, Bob Abraham, for his
 thorough reading of the manuscript and positive feedback. We would
 like to thank Pieter van Dokkum for helpful suggestions and Sandy
 Faber for stimulating discussion.  A. v. d. W. acknowledges support
 from NASA grant NAG5-7697.}

\bibliographystyle{apj}

\end{document}

%% file: tab1.tex
\begin{deluxetable*}{cccc}
\tabletypesize{\scriptsize}
\tablecolumns{4}
\tablewidth{0pt}
\tablenum{1}
\tablecaption{Magnitude and Color Conversions}
\startdata
\hline
\hline
$z$    & $g_0$                     & $(u-g)_0$             & $(g-r)_0$             \\
\hline
$0.02$ & $g - 0.096 (g-r) + 0.019$ & $1.035 (u-g) - 0.111$ & $0.951 (g-r) - 0.001$ \\
$0.03$ & $g - 0.141 (g-r) + 0.027$ & $1.037 (u-g) - 0.124$ & $0.928 (g-r) - 0.003$ \\
$0.04$ & $g - 0.185 (g-r) + 0.035$ & $1.029 (u-g) - 0.123$ & $0.906 (g-r) - 0.005$ \\
\hline
$z$    & $B_0$                     & $(U-B)_0$             & $(B-V)_0$             \\
\hline
$0.60$ & $z + 0.559 (v-z) + 0.428$ & $0.732 (v-i) - 0.703$ & $0.501 (v-z) - 0.049$ \\
$0.70$ & $z + 0.413 (v-z) + 0.511$ & $0.696 (v-i) - 0.757$ & $0.463 (v-z) - 0.075$ \\
$0.80$ & $z + 0.289 (v-z) + 0.617$ & $0.661 (v-i) - 0.778$ & $0.411 (v-z) - 0.082$ \\
$0.90$ & $z + 0.466 (i-z) + 0.792$ & $0.404 (v-z) - 0.666$ & $0.982 (i-z) + 0.030$ \\
$1.00$ & $z + 0.194 (i-z) + 0.847$ & $0.371 (v-z) - 0.628$ & $1.044 (i-z) - 0.142$
\enddata

\tablecomments{ Transformations used to calculate rest-frame
magnitudes and colors from observed magnitudes and colors for eight
different redshifts.  The letters $v$, $i$, and $z$ are shorthand for
$v_{606}$, $i_{775}$, and $z_{850}$, respectively. The magnitudes are
total magnitudes for which the definitions can be found in
\S\S~\ref{secsdss} and \ref{secdatacdfs}. The total magnitudes $g_0$
and $B_0$ are not corrected for cosmological surface brightness
dimming. Note that $B_0$, $(U-B)_0$, and $(B-V)_0$ are on the Vega
system, whereas all other magnitudes and colors are on the AB
system. The $u-g$ conversion contains a correction ($-0.04$ mag) to
account for the difference between the SDSS $u$-band zero point and
the true AB zero point.}
\label{tab1}
\end{deluxetable*}

%% file: tab2.tex
\begin{deluxetable*}{cccccccccc}
\tabletypesize{\scriptsize}
\tablecolumns{10}
\tablewidth{0pt}
\tablenum{2}
\tablecaption{Sample Characteristics}
\tablehead {
\colhead{} &
\colhead{} &
\colhead{$f_{\rm{ET}}$} &
\colhead{$f_{\rm{RS}}$} &
\colhead{$f_{\rm{n}>2.5}$} &
\colhead{Med$(\Sigma_{\rm{M}})$} &
\colhead{Med$(\Sigma_{\rm{L}})$} &
\colhead{} &
\colhead{$<\rm{SFR}>$} &
\colhead{Med(SFR)} \\
\colhead{$z$} &
\colhead{$N$} &
\colhead{(\%)} &
\colhead{(\%)} &
\colhead{(\%)} &
\colhead{Mpc$^{-2}$} &
\colhead{Mpc$^{-2}$} &
\colhead{Med$(U-B)$} &
\colhead{$M_{\odot}~\rm{yr}$} &
\colhead{$M_{\odot}~\rm{yr}$}\\
\colhead{(1)} &
\colhead{(2)} &
\colhead{(3)} &
\colhead{(4)} &
\colhead{(5)} &
\colhead{(6)} &
\colhead{(7)} &
\colhead{(8)} &
\colhead{(9)} &
\colhead{(10)}}
\startdata
$0.020 < z < 0.045$ & 2003 & $43\pm3$ & $68\pm1$ & $64\pm2$ &  0.17 & 0.4 & 0.40 &  $3.1\pm1.5$ & $0.4\pm0.2$ \\
$0.6   < z < 1.0  $ &  207 & $48\pm7$ & $64\pm3$ & $72\pm4$ &   1.0 & 2.1 & 0.23 &  $12\pm6$    & $\lesssim 3$ \\
\enddata

\tablecomments{ Col. (1) Redshift. Col. (2) Number of galaxies in the
  sample. Col (3) Early-type galaxy fraction as inferred from the
  automated $B-n$ classifications (see \S~\ref{secmorph}). Col. (4)
  Fraction of galaxies located on the red sequence as defined in
  \S~\ref{seccolormass}. Col. (5) Fraction of galaxies with Sersic $n$
  parameter larger than $n=2.5$ (\S~\ref{secmorph}). Col. (6) Median
  local surface density of galaxies with masses $M>4\times
  10^{10}~M_{\odot}$ (see \S~\ref{secdens}). Col. (7) Median local
  surface density of galaxies with luminosities
  $M_{\rm{V}}+0.8z<-19.78$ (see \S~\ref{secdens}). Col. (8) Median
  rest-frame $U-B$ color as computed in \S\S~\ref{secsdss} and
  \ref{secdatacdfs}. Col. (9) Mean SFR (\S~\ref{secsf}); (9) Median
  SFR (\S~\ref{secsf}). The errors on the SFRs are assumed to be
  50\%, mainly to account for systematic uncertainties.}
\label{tab2}
\end{deluxetable*}